\documentclass[12pt]{iopart}

\usepackage{iopams}

\usepackage{amssymb}
\usepackage{graphicx}
\usepackage{multirow}
\usepackage{bm}
\usepackage{url}
\usepackage{cite}

\newtheorem{definition}{Definition}
\newtheorem{proposition}[definition]{Proposition}
\newtheorem{lemma}[definition]{Lemma}

\newtheorem{theorem}[definition]{Theorem}
\newtheorem{corollary}[definition]{Corollary}

\begin{document}

\title{SIC~POVMs and Clifford groups in prime dimensions}
\author{Huangjun Zhu}
\address{Centre for Quantum Technologies, %
National University of Singapore, Singapore 117543, Singapore}
\address{NUS Graduate School for Integrative Sciences and
Engineering, Singapore 117597, Singapore}
\ead{zhuhuangjun@nus.edu.sg}

\begin{abstract}
We show that in prime dimensions not equal to three, each group
covariant symmetric informationally complete positive operator
valued measure (SIC~POVM) is covariant with respect to  a unique
Heisenberg--Weyl (HW) group. Moreover,  the symmetry group of the
SIC~POVM is a subgroup of the Clifford group. Hence, two SIC~POVMs
covariant with respect to the HW group are unitarily or
antiunitarily equivalent if and only if they are on the same orbit
of the extended Clifford group. In dimension three, each group
covariant SIC~POVM may be covariant with respect to three or nine HW
groups, and the symmetry group of the SIC~POVM is a subgroup of at
least one of the Clifford groups of these HW groups respectively.
There may exist two or three orbits of equivalent SIC~POVMs for each
group covariant SIC~POVM, depending on the order of its symmetry
group. We then establish a complete equivalence relation among group
covariant SIC~POVMs in dimension three, and classify inequivalent
ones according to the geometric phases associated with fiducial
vectors. Finally, we uncover additional SIC~POVMs by regrouping of
the fiducial vectors from different SIC~POVMs which may or may not
be on the same orbit of the extended Clifford group.
\end{abstract}



\date{draft of \today}

\maketitle

\tableofcontents

\section{Introduction}
A positive operator valued measure (POVM) is the most general
measurement in quantum theory. It consists of a set of outcomes
represented mathematically as a set of positive operators ${\Pi_j}$s
satisfying $\sum_j\Pi_j=I$, where $I$ is the identity. Given an
input quantum state $\rho$, the probability of obtaining the outcome
$\Pi_j$ is given by $p_j=\mathrm{tr}(\rho\Pi_j)$. An
\emph{informationally complete} POVM (IC POVM)  is one  which allows
us to  reconstruct any quantum state according to the set of
probabilities $p_j$s. Simple parameter counting shows that an IC
POVM contains at least $d^2$ outcomes for  a $d$-dimensional Hilbert
space. An IC POVM with $d^2$ outcomes is called \emph{minimal}.

A \emph{symmetric informationally complete} POVM (SIC~POVM)
\cite{Zau99, RBSC04, App05, SG10} is a special minimal IC POVM which
consists of $d^2$ pure  subnormalized projectors with equal pairwise
fidelity. It is considered as a fiducial POVM due to its high
symmetry and high tomographic efficiency \cite{Fuc02,RBSC04,App05,
Sco06, ZTE10}. The existence of SIC~POVMs in any finite dimension
was first conjectured by Zauner \cite{Zau99} about ten years ago,
and has been confirmed numerically in dimensions up to $67$
\cite{RBSC04, SG10}. Analytical solutions have been found in
dimensions 2, 3 \cite{DGS75}; 4, 5 \cite{Zau99}; 6 \cite{Gra04}; 7
\cite{App05}; 8 \cite{Hog98, Gra05}; 9, $\ldots$, 15 \cite{SG10,
Gra05,Gra06,Gra08a,Gra08b}; 19 \cite{App05}; 24, 35, 48 \cite{SG10}.
It is generally believed that SIC~POVMs  exist in any finite
dimension, however, a rigorous mathematical proof is not known.

In addition to their application in quantum state tomography,
SIC~POVMs are also interesting for many other reasons. They are
closely related to mutually unbiased bases (MUB) \cite{Woo04, ADF07,
App08}. They are studied under the name of equiangular lines
\cite{LS73} in the mathematical community, and  are well known as
minimal 2-design in design theory \cite{RBSC04}. The Lie algebraic
significance of SIC~POVMs  was also discussed recently \cite{AFF09}.

A \emph{group covariant} SIC~POVM is one which can be generated from
a single vector---\emph{fiducial vector}---under the action of a
group consisting of unitary operations. Almost all known SIC~POVMs
are covariant with respect to the \emph{Heisenberg--Weyl (HW) group}
or generalized Pauli group. The \emph{Clifford group} is the
normalizer of the HW group which consists of unitary operations, and
the \emph{extended Clifford group} is the larger group which
contains also antiunitary operations \cite{App05, App09}. Obviously,
a fiducial vector remains a fiducial vector when transformed by any
element in the extended Clifford group. Fiducial vectors and
SIC~POVMs  form disjoint orbits under the action of the (extended)
Clifford group. SIC~POVMs on the same orbit of the extended Clifford
group are \emph{unitarily or antiunitarily equivalent} in the sense
that they can be transformed into each other with unitary or
antiunitary operations.

Except in a few small dimensions, there are generally more than one
orbits of SIC~POVMs according to the numerical searches performed by
Scott and Grassl \cite{SG10}. Hence, a natural question arise: Are
two SIC~POVMs on two different orbits of the (extended) Clifford
group equivalent? This question is closely related to the following
open question: For an HW covariant SIC~POVM, is its (extended)
symmetry group a subgroup of the (extended) Clifford group? By
 \emph{symmetry group} (\emph{extended symmetry group}) of a SIC~POVM, we mean the set of all
 unitary (unitary or antiunitary) operations which leave
the SIC~POVM invariant.  Although an affirmative answer to the later
question is tacitly assumed by many researchers in the community, a
rigorous proof is yet unavailable in the literature. In this paper
we settle to answer all these questions for prime dimensions.

Dimension three is the only known case where there exist continuous
orbits of SIC~POVMs. Despite the low dimension, and the fact that
fiducial vectors have been known for a long time \cite{Zau99,
RBSC04, App05}, a complete picture of SIC~POVMs in dimension three
has yet to be uncovered. We shall show that there are some
additional peculiarities about SIC~POVMs in dimension three in
contrast with the ones in other prime dimensions. Moreover, we shall
provide new insights about SIC~POVMs in dimension three by
establishing a complete equivalence relation among all group
covariant SIC~POVMs and classifying all inequivalent ones according
to the \emph{geometric phases} \cite{Ber84, AA87} or Bargmann
invariants \cite{Bar64} associated with fiducial vectors. In
addition, we uncover additional SIC~POVMs through regrouping of the
fiducial vectors from different SIC~POVMs which may or may not be on
the same orbit of the extended Clifford group.

The  paper is organized as follows. In \sref{sec:pre}, we recall the
basic properties  of  the SIC~POVMs and Clifford groups in prime
dimensions. In \sref{sec:main}, we prove  that, in any prime
dimension, each group covariant SIC~POVM is covariant with respect
to the HW group, and the symmetry group of the SIC~POVM is a
subgroup of certain Clifford group. The implications of these
results on the equivalence relation among group covariant SIC~POVMs
are also discussed in detail. In \sref{sec:sicpom3}, we establish a
complete equivalence relation among all group covariant SIC~POVMs in
dimension three, and classify  inequivalent ones. We also uncover
additional SIC~POVMs by regrouping of the fiducial vectors. We
conclude with a summary.

\section{\label{sec:pre}Preliminary about SIC~POVMs and Clifford groups in prime
dimensions}
\subsection{\label{sec:SICHW}SIC~POVMs and  Heisenberg--Weyl group}
In a $d$-dimensional Hilbert space, a SIC~POVM \cite{Zau99, RBSC04,
App05} consists of $d^2$ outcomes that are  subnormalized projectors
onto pure states $\Pi_j=\frac{1}{d}|\psi_j\rangle\langle\psi_j|$ for
$j=1,\ldots,d^2$, such that
\begin{eqnarray}\label{eq:SIC}
|\langle\psi_j|\psi_k\rangle|^2=\frac{1+d\delta_{jk}}{d+1}.
\end{eqnarray}
The  symmetry group $G_{\mathrm{sym}}$  of a SIC~POVM consists of
all unimodular  unitary  operators (unimodular operators or matrices
are those with determinant 1) that leave the SIC~POVM invariant,
that is, permute the set of vectors $|\psi_j\rangle$s up to some
phase factors. Since operators which differ only by  overall phase
factors implement essentially the same transformation,
 they can be identified into equivalence classes. Let $G$ be any unitary group, and $U$ any element in
 $G$.
Throughout this paper, $\bar{U}$ is used to denote the equivalence
class of $U$, and the expression  $U^\prime \in \bar{U}$
 means
 that $U^\prime $ is in the equivalence class of $U$. With the
 product rule $\bar{U}\bar{U}^\prime=\overline{UU^\prime}$, the set of $\bar{U}$s
 form  the \emph{collineation group} $\bar{G}$ of $G$
 \cite{Bli917}. Let $\Phi(G)$ be the subgroup of $G$ consisting of
 elements which are proportional to the identity; then $\bar{G}$ is
 the quotient group of $G$ with respect to $\Phi(G)$, that is,
 $\bar{G}=G/\Phi(G)$. There  exist infinitely many different unitary groups $G$ with  the same
 collineation group $\bar{G}$.  However, if G is  unimodular, then  $|G|\leq
 d|\bar{G}|$, where $|G|$ ($|\bar{G}|$) denotes the order of $G$ ($|\bar{G}|$). Moreover, there is a unique unimodular unitary group
 $G^\prime$ that satisfies  $\bar{G^\prime}=\bar{G}$ and $|G^\prime|=d|\bar{G}|$  \cite{Bli917}. When there is no confusion, $G$ and $\bar{G}$
 will be referred to with the same name. For example
 $\bar{G}_{\mathrm{sym}}$ is also  called the symmetry group of the
 SIC~POVM, and it is generally more convenient to work with $\bar{G}_{\mathrm{sym}}$ rather than
 $G_{\mathrm{sym}}$. We shall
some times work with unimodular unitary groups and some times with
collineation groups, depending on which one is more convenient. To
simplify the notation, we shall often
 denote an element in the collineation group with a representative which need
not be unimodular.

The group $\bar{G}_{\mathrm{sym}}$ can also be defined in an
alternative way. Let  $G_{\mathrm{sym}}^\prime$  be the group
consisting of  all unitary operations that leave the SIC~POVM
invariant, then
$\bar{G}_{\mathrm{sym}}=\bar{G}_{\mathrm{sym}}^\prime=G_{\mathrm{sym}}^\prime/\Phi(G_{\mathrm{sym}}^\prime)$.
The group $G_{\mathrm{sym}}^\prime$ is also  called  the symmetry
group of the SIC~POVM. The advantage of the second definition is
that it can be extended to cover antiunitary operations
unambiguously. The extended symmetry group
$EG_{\mathrm{sym}}^\prime$ of a SIC~POVM consists of all unitary or
antiunitary operations that leave the SIC~POVM invariant. The
quotient group
$\overline{EG}_{\mathrm{sym}}^\prime=EG_{\mathrm{sym}}^\prime/\Phi(EG_{\mathrm{sym}}^\prime)$
is also called the extended symmetry group. To impose the unimodular
constraint on the symmetry group is mainly to ensure that the group
$G_{\mathrm{sym}}$ be finite, which is crucial in later discussions.
In the rest of the paper, concerning the symmetry group of a
SIC~POVM, we only consider the collineation group and unimodular
unitary group; concerning the extended symmetry group, we only
consider the collineation group, and we write
$\overline{EG}_{\mathrm{sym}}$ in place of
$\overline{EG}_{\mathrm{sym}}^\prime$ to simplify the notation.

Since a SIC~POVM is informationally complete, any unitary operator
that
 stabilizes all fiducial vectors must be proportional to the identity. Hence, the action of  $\bar{G}_{\mathrm{sym}}$ on the set
 of vectors in the SIC~POVM is faithful, which implies that  $\bar{G}_{\mathrm{sym}}$ is isomorphic to a subgroup of the
 full symmetry group of $d^2$ letters, and is thus a finite group. As a result,
 $G_{\mathrm{sym}}$ is also a finite group due to the relation $|G|\leq
 d|\bar{G}|$.

Under the action of $\bar{G}_{\mathrm{sym}}$, the vectors in the
SIC~POVM form disjoint orbits. The \emph{stability group} or
stabilizer of a vector  in the SIC~POVM is the group consisting of
all operations that leave the vector invariant. A SIC~POVM is
 group covariant if the vectors of the SIC~POVM form a single
orbit under the action of $\bar{G}_{\mathrm{sym}}$. In that case,
the SIC~POVM is covariant with respect to $G_{\mathrm{sym}}$ or
$\bar{G}_{\mathrm{sym}}$, and each vector in the SIC~POVM is a
fiducial vector.  More generally, if the fiducial vectors of a group
covariant SIC~POVM are on the same orbit of a subgroup $\bar{G}$  of
$\bar{G}_{\mathrm{sym}}$, then the SIC~POVM is  covariant with
respect to $G$ or $\bar{G}$. For a group covariant SIC~POVM, the
stability group of each fiducial vector is conjugated to each other,
and $|\bar{G}_{\mathrm{sym}}|/|\bar{K}|=d^2$, where  $\bar{K}$ is
the stability group of any fiducial vector.

Two SIC~POVMs are unitarily or antiunitarily  equivalent if there
exists a unitary or antiunitary operator that maps the vectors of
one SIC~POVM to that of the other one up to a permutation of the
vectors in addition to  some phase factors.

Most known SIC~POVMs are covariant with respect to the
Heisenberg--Weyl (HW) group or generalized Pauli group $D$
\cite{Zau99, RBSC04, App05}, which is generated by the two operators
$X, Z$ defined below,
\begin{eqnarray}
Z|e_r\rangle&=&\omega^r|e_r\rangle, \nonumber\\
X|e_r\rangle&=&\left\{ \begin{array}{cl}
  |e_{r+1}\rangle & r=0,1,\ldots,d-2, \nonumber\\
  |e_0\rangle & r=d-1, \\
\end{array}\right.\\
D_{k_1,k_2}&=&\tau^{k_1 k_2}X^{k_1}Z^{k_2},\label{HW}
\end{eqnarray}
where $\omega=\rme^{2\pi \rmi/d}$, $\tau=-\rme^{\pi \rmi/d}$,
$k_1,k_2\in Z_d$, and $Z_d$ is the additive group of integer modulo
$d$. The group $D$ consists of  $d^3$ elements:
$\omega^{k_3}D_{k_1,k_2}$ for $k_1,k_2,k_3\in Z_d$, while $\bar{D}$
consists of $d^2$ elements: $D_{k_1,k_2}$  for $k_1,k_2\in Z_d$
(here we denote the elements in the collineation group with the
representatives in the linear group). In addition, $\bar{D}$ is
abelian, while $D$ is not. Due to the commutation relation
$XZX^{-1}Z^{-1}=\omega^{-1}I$, any unitary group with $\bar{D}$ as
its collineation group must contain the subgroup generated by
$\omega I$. Moreover, there exists a unique unimodular unitary group
with $\bar{D}$ as its collineation group, this unique unimodular
unitary group is also referred to as the HW group. When $d$ is odd,
which is the most relevant case in this paper, the group $D$ defined
in \eref{HW} is already unimodular. When $d$ is even, it can be made
unimodular by some phase factors. For example, in dimension two, the
unimodular form of the HW group consists of the following eight
elements $\pm I, \pm iX,\pm iZ, \pm XZ$.

A fiducial vector $|\psi\rangle$ of the HW group  satisfies the
following equation \cite{Zau99, RBSC04, App05},
\begin{eqnarray}
|\langle\psi|D_{\mathbf{k}}|\psi\rangle|=\frac{1}{\sqrt{d+1}}\quad
\mbox{for}\quad \mathbf{k}\neq 0,
\end{eqnarray}
where $\mathbf{k}=(k_1,k_2)^T$.

The Clifford group $\mathrm{C}(d)$ is  the normalizer of the HW
group which consists of unimodular unitary operators; its
collineation group $\bar{\mathrm{C}}(d)$ is  also called the
Clifford group. As in the case of HW group,
$|\mathrm{C}(d)|=d|\bar{\mathrm{C}}(d)|$. The extended Clifford
group $\overline{\mathrm{EC}}(d)$ is the larger group which contains
 also antiunitary operators, and it is generated by $\bar{\mathrm{C}}(d)$
and the complex conjugation operator $\hat{J}$ \cite{App05, App09}.
By definition, for any element $U$ in the extended Clifford group,
$U|\psi\rangle$ is  a fiducial vector whenever $|\psi\rangle$ is
\cite{App05}. Fiducial vectors and SIC~POVMs form  disjoint orbits
under the action of the (extended) Clifford group. SIC~POVMs on the
same orbit of the extended Clifford group are unitarily or
antiunitarily equivalent. For SIC~POVMs on different orbits, there
is no simple criterion so far for determining their equivalence
relation. In this paper we shall solve this problem for prime
dimensions.

In the rest of the paper, except when stated otherwise, we assume
that the dimension of the Hilbert space is a  prime $p$, and we are
only concerned with group covariant SIC~POVMs. In prime dimensions,
for any unimodular unitary group $G$, either $|G|=|\bar{G}|$ or
$|G|=p|\bar{G}|$ is satisfied \cite{Bli917}. Most unimodular unitary
groups we shall consider contain the HW group as a subgroup and thus
contain also the subgroup generated by $\omega I$; hence
$|G|=p|\bar{G}|$. As a consequence, there is a one-to-one
correspondence between unimodular unitary groups  and their
collineation groups.

It turns out that, in any prime dimension, a group covariant
SIC~POVM must be covariant with respect to the HW group (see
\sref{sec:HWcovariance}). Thus the Clifford group plays a crucial
role in classifying group covariant SIC~POVMs.  In the rest of this
section we focus on the Clifford groups in \emph{odd prime}
dimensions (see also \cite{App05, App09}), in preparation for the
discussion in the next section. Some results presented here may also
be of independent interests. The Clifford group in dimension two
will be discussed briefly in \sref{sec:d2}. Before discussing the
Clifford group, we need to take a detour reviewing the properties of
the special linear group $\mathrm{SL}(2,p)$ for odd prime $p$.

\subsection{\label{sec:SL}Special linear group $\mathrm{SL}(2,p)$}
The special linear group  $\mathrm{SL}(2,p)$ over the field $Z_p$
consists of $2\times 2$ matrices with entries from $Z_p$, and unit
determinant mod $p$, that is elements of the form
\begin{eqnarray}\label{eq:F}
F=\left(%
\begin{array}{cc}
  \alpha & \beta\\
  \gamma & \delta \\
\end{array}%
\right),
\end{eqnarray}
where $\alpha\delta-\beta\gamma=1$ mod $p$. Likewise, the extended
special linear group  $\mathrm{ESL}(2,p)$ is the larger group which
contains also  $2\times 2$ matrices with determinant $-1$ mod $p$.
The orders of $\mathrm{SL}(2,p)$ and $\mathrm{ESL}(2,p)$ are
$p(p^2-1)$ and $2p(p^2-1)$ respectively. Since $\mathrm{ESL}(2,p)$
is a union of two cosets, that is
$\mathrm{ESL}(2,p)=\mathrm{SL}(2,p)\cup J\mathrm{SL}(2,p)$, where
\begin{eqnarray}
J=\left(
    \begin{array}{cc}
      1 & 0 \\
      0 & -1 \\
    \end{array}
  \right),
\end{eqnarray}
it is enough to focus on  $\mathrm{SL}(2,p)$ in the following
discussion.

The centre of the special linear group $\mathrm{SL}(2,p)$ is
generated by $\mathrm{diag}(-1,-1)$,  the unique order 2 element in
the group. The quotient group of $\mathrm{SL}(2,p)$ with respect to
its centre---the projective special linear group
$\mathrm{PSL}(2,p)$---is a simple group (a group without nontrivial
normal subgroups) for $p\geq5$ \cite{Dic58, KS04}.

There are $p+4$ conjugacy classes in $\mathrm{SL}(2,p)$ \cite{Hum75,
ABC07, App09}. \Tref{tab:class1} shows the class representatives,
their orders and numbers of conjugates determined by Humphreys
\cite{Hum75}, where

\begin{table}
\centering     \caption{Class representatives, and their numbers of
conjugates  in $\mathrm{SL}(2,p)$ for odd prime $p$ from Humphreys
\cite{Hum75},  orders of these class representatives are also
included for completeness. Here the class representatives $\bm{1},
z, c_1,c_2, a$ are defined in \eref{eq:SLrep} (the class
representatives $c_1, c_2$ are modified for convenience),
 $b$ is an element of order
$p+1$, $1\leq l\leq \frac{p-3}{2}, 1\leq m\leq \frac{p-1}{2}$ and
$\mathrm{gcd}(l,p-1)$ denotes the greatest common divisor of $l$ and
$p-1$, similarly for $\mathrm{gcd}(m,p+1)$.}\label{tab:class1}
\begin{math}
\begin{array}{ccccccccc}
\br \mbox{Representative}& \bm{1} & z & a^l & b^m & c_1 & c_2 & zc_1
& zc_2
\\[0.5ex]
 \hline
\mbox{Order} & 1 & 2& \frac{p-1}{\mathrm{gcd}(l,p-1)} &
\frac{p+1}{\mathrm{gcd}(m,p+1)} & p & p & 2p&
2p\\[1ex]
\begin{tabular}{c}
  \mbox{Number of}  \\
  \mbox{conjugates}  \\
\end{tabular}& 1 & 1 & p(p+1) & p(p-1) & \frac{1}{2}(p^2-1) & \frac{1}{2}(p^2-1) & \frac{1}{2}(p^2-1)
  &\frac{1}{2}(p^2-1)\\
  \br
\end{array}
\end{math}

\end{table}
\begin{eqnarray}
\fl\label{eq:SLrep} \bm{1}&=&\left(
    \begin{array}{cc}
      1 & 0 \\
      0 & 1 \\
    \end{array}
  \right),\qquad
  z=-\bm{1},\qquad
  c_1=\left(
    \begin{array}{cc}
      1 & 0 \\
      1 & 1 \\
    \end{array}
  \right),\qquad
  c_2=\left(
    \begin{array}{cc}
      1 & 0 \\
      \nu & 1 \\
    \end{array}
  \right),\qquad 
    a=\left(
    \begin{array}{cc}
      \nu & 0 \\
      0& \nu \\
    \end{array}
  \right)
\end{eqnarray}
and $\nu$ is a primitive element in $Z_p$.  For the class
representative $c_2$ or $zc_2$, $\nu$ can also be any element in
$Z_p$ that is not a quadratic residue, that is not a square of any
element in $Z_p$. There is one class (class representative $z$,
actually only one element)  with elements of order 2, two classes
(class representatives $c_1, c_2$) with elements of order $p$, and
two classes (class representatives $zc_1, zc_2$) with elements of
order $2p$. For each divisor $k$ of $p-1$ which is not equal to 2,
the number of classes with elements of order $k$ is equal to
$\frac{1}{2}\varphi(k)$. Here
 $\varphi(k)$ is the number of elements of
order $k$ in any cyclic group whose  order is divisible by $k$. It
is also known as the Euler function which denotes the number of
positive integers which are less than $k$  and coprime with $k$.
Similarly, for each divisor $k$ of $p+1$ which is not equal to 2,
the number of classes with elements of order $k$ is
$\frac{1}{2}\varphi(k)$.

There are $p+1$ Sylow $p$-subgroups in $\mathrm{SL}(2, p)$, $Q_1,
\ldots, Q_{p+1}$, and all of them are conjugated to each other
according to Sylow's theorem \cite{KS04}. The normalizer $N_j$s of
$Q_j$s are also conjugated to each other. Suppose $Q_1$ consists of
the following elements:
\begin{eqnarray}
\label{eq:Q1}
 \left(
  \begin{array}{cc}
    1 & 0 \\
    \gamma & 1 \\
  \end{array}
\right) \qquad \mbox{for}\quad \gamma\in Z_p,
\end{eqnarray}
then the normalizer $N_1$ of $Q_1$ consists of the following
elements:
\begin{eqnarray}
\label{eq:N1} \left(
  \begin{array}{cc}
    \alpha & 0 \\
    \gamma & \alpha^{-1} \\
  \end{array}
\right) \qquad \mbox{for}\quad \gamma\in Z_p,\quad \alpha\in Z_p^*,
\end{eqnarray}
where $Z_p^*$ is the multiplicative group consisting of nonzero
elements in $Z_p$, which is also cyclic \cite{KS04}. The order of
$N_1$ is  $p(p-1)$; it is  cyclic if $p=3$ and  not cyclic if
$p\geq5$. In addition, each subgroup of $N_1$ whose order is equal
to $2p$ or is not a multiple of $p$ is cyclic.


$\mathrm{SL}(2,p)\ltimes (Z_p)^2$ is the semidirect product group of
$\mathrm{SL}(2,p)$ and $(Z_p)^2$ equipped with the following product
rule:
\begin{eqnarray}
\label{eq:productrule}
(F_1,\chi_1)\circ(F_2,\chi_2)=(F_1F_2,\chi_1+F_1\chi_2),
\end{eqnarray}
where $F_1,F_2\in \mathrm{SL}(2, p)$, and $\chi_1, \chi_2\in
(Z_p)^2$. Similarly, $\mathrm{ESL}(2,p)\ltimes (Z_p)^2$ is defined
with the same product rule. The orders of $\mathrm{SL}(2,p)\ltimes
(Z_p)^2$ and $\mathrm{ESL}(2,p)\ltimes (Z_p)^2$ are $p^3(p-1)$ and
$2p^3(p-1)$ respectively \cite{App05}. In the following discussion,
we focus on $\mathrm{SL}(2,p)\ltimes (Z_p)^2$.

The conjugacy classes of $\mathrm{SL}(2,p)\ltimes (Z_p)^2$ can be
determined based on the conjugacy classes of $\mathrm{SL}(2,p)$, see
also \cite{ABC07, App09}. Since elements of the form $(\bm{1},\chi)$
for $\chi\neq\bm{0}$ form a single conjugacy class, it remains to
consider $(F,\chi)$ with $F\neq \bm{1}$. Due to the product rule in
\eref{eq:productrule}, it suffices to deal with the case where $F$
is a class representative of $\mathrm{SL}(2,p)$ listed in
\tref{tab:class1}. Consider the following equality:
\begin{eqnarray}
\label{eq:product}
(1,\chi_1)\circ(F,\chi)\circ(1,\chi_1)^{-1}=(F,(1-F)\chi_1+\chi);
\end{eqnarray}
if $F\neq c_1, c_2$, then $1-F$ is nonsingular, and we can choose
$\chi_1=-(1-F)^{-1}\chi$ to  eliminate the term $(1-F)\chi_1+\chi$;
hence $(F,\chi)$ is conjugated to $(F,\bm{0})$.  If $F=\Bigl(
                                                            \begin{array}{cc}
                                                              1 & 0 \\
                                                             \gamma & 1 \\
                                                            \end{array}
                                                          \Bigr)
$,   according to a similar argument, $\Bigl(F,\Bigl(
                                                           \begin{array}{c}
                                                             k_1 \\
                                                             k_2 \\
                                                           \end{array}
                                                         \Bigr)\Bigr)
$ is conjugated to $\Bigl(F,\Bigl(
\begin{array}{c}
k_1 \\
 0 \\
\end{array}
\Bigr)\Bigr)$; in addition,  $\Bigl(F,\Bigl(
\begin{array}{c}
k_1 \\
 0 \\
\end{array}
\Bigr)\Bigr)$ is conjugated to $\Bigl(F,\Bigl(
\begin{array}{c}
k_1^\prime \\
 0 \\
\end{array}
\Bigr)\Bigr)$ if and only if $k_1^\prime=k_1$ or $k_1^\prime=-k_1$.

\begin{table}
  \centering
  \caption{\label{tab:class2}   Class representatives, their orders  and numbers of conjugates  in
  $\mathrm{SL}(2,p)\ltimes(Z_p)^2$ for odd prime $p$.
  Here $\bm{1}, z, c_1,c_2, a$  are defined in \eref{eq:SLrep}, $b$ is an element of order $p+1$ in $\mathrm{SL}(2,p)$, $1\leq
l\leq\frac{p-3}{2}$, $1\leq m\leq\frac{p-1}{2}$, $1\leq
k_1,k_2\leq\frac{p-1}{2}$  and $\mathrm{gcd}(l,p-1)$ denotes the
greatest common divisor of $l$ and $p-1$, similarly for
$\mathrm{gcd}(m,p+1)$.}
\begin{math}
\begin{array}{cccccccc}
\br \mbox{Representative} & (\bm{1},\bm{0})&
\Bigl(\bm{1},\Bigl(\begin{array}{c}
                     1\\
                    0
                  \end{array}
 \Bigr)\Bigr) & (z,\bm{0}) & (a^l,\bm{0}) & (b^m,\bm{0}) & (c_1,\bm{0})
  \\ \hline
\mbox{Order} &1 &p & 2 & \frac{p-1}{\mathrm{gcd}(l,p-1)} &
\frac{p+1}{\mathrm{gcd}(m,p+1)} & p
 \\
\begin{tabular}{c}
  \mbox{Number of}  \\
  \mbox{conjugates}  \\
\end{tabular}
& 1 &p^2-1& p^2 & p^3(p+1) & p^3(p-1) & \frac{1}{2}p(p^2-1)
 \\
  \br
\mbox{Representative}& \Bigl(c_1,\Bigl(\begin{array}{c}
                     k_1\\
                    0
                  \end{array}
 \Bigr)\Bigr)&
(c_2,\bm{0}) &
  \Bigl(c_2,\Bigl(\begin{array}{c}
                     k_2\\
                    0
                  \end{array}
 \Bigr)\Bigr)&(zc_1,\bm{0}) &
  (zc_2,\bm{0})  \\
  \hline
  \mbox{Order}&p &  p &p &2p & 2p\\
 \begin{tabular}{c}
  \mbox{Number of}  \\
  \mbox{conjugates}  \\
\end{tabular}&p(p^2-1)& \frac{1}{2}p(p^2-1) & p(p^2-1)& \frac{1}{2}p^2(p^2-1)
  &\frac{1}{2}p^2(p^2-1)\\
  \br
\end{array}
\end{math}
\end{table}
The class representatives, their orders, and numbers of conjugates
in $\mathrm{SL}(2,p)\ltimes (Z_p)^2$ are shown in \tref{tab:class2}.
There are $2p+4$ conjugacy classes, $p+2$ of which consist of
elements of order $p$.
 The number of classes with elements of any other order is the same as that in
 $\mathrm{SL}(2,p)$. That is, one class of order 2, two classes of order $2p$,
 and $\frac{1}{2}\varphi(k)$ classes of order $k$ for each divisor $k$ of $p-1$ or
 $p+1$ which is not equal to 2, where $\varphi(k)$ is the Euler
 function. Note that $\varphi(k)$ is equal to 0, 1,
 2, 2, 4, 2 respectively for $k=1,\ldots, 6$,
 and $\varphi(k)>2$  for any other positive integer $k$. It follows that,  if $p>3$, all order 3 elements are conjugated to
 each other, so are all order 4 elements and order 6 elements, and
 there are more than one classes   if  the order is a divisor of $p-1$ or $p+1$ which is equal to 5 or
 larger than 6.

\subsection{\label{sec:Clifford}Clifford group}

 An important step towards understanding the structures of
the (extended) Clifford groups and SIC~POVMs  is the following
isomorphism given by Appleby \cite{App05},
\begin{eqnarray}
f_E&:&  \mathrm{ESL}(p)\ltimes (Z_p)^2\rightarrow
\overline{\mathrm{EC}}(p), \nonumber\\
&&UD_{\mathbf{k}}U^\dag=\omega^{\langle
\chi,F\mathbf{k}\rangle}D_{F\mathbf{k}}\qquad \mbox{for} \quad
U=f_E(F,\chi),\label{eq:isomorphism1}
\end{eqnarray}
where $\langle \mathbf{k},\mathbf{q}\rangle=k_2q_1-k_1q_2$. Here is
the explicit correspondence if $\mathrm{det}(F)=1$ (assuming $F$ is
given in \eref{eq:F}) and  $\beta\neq0$,
\begin{eqnarray}
&(F,\chi)\rightarrow U=D_\chi V_F,&\nonumber\\
&V_F=\frac{1}{\sqrt{p}}\sum\limits
_{r,s=0}^{p-1}\tau^{\beta^{-1}(\alpha s^2-2rs+\delta
r^2)}|e_r\rangle\langle e_s|.&\label{eq:VF1}
\end{eqnarray}
If $\beta=0$, then $\alpha, \delta\neq0$, $\alpha\delta=1$, and $F$
can be written as the product of the following two matrices:
\begin{eqnarray}
F_1=\left(
  \begin{array}{cc}
    0 & -1 \\
    1 & 0 \\
  \end{array}
\right)\qquad \mbox{and}\qquad F_2=\left(
  \begin{array}{cc}
    \gamma & \delta \\
     -\alpha & 0 \\
  \end{array}
\right),
\end{eqnarray}
such that $V_{F_1}$ and $V_{F_2}$ can be computed according to
\eref{eq:VF1}. Hence we have $(F,\chi)\rightarrow D_\chi
V_{F}=D_\chi V_{F_1}V_{F_2}$, where
\begin{eqnarray}
\label{eq:VF2} V_{F_1}&=&\frac{1}{\sqrt{p}}\sum\limits
_{r,s=0}^{p-1}\tau^{2rs}|e_r\rangle\langle e_s|,\nonumber\\
V_{F_2}&=&\frac{1}{\sqrt{p}}\sum\limits
_{r,s=0}^{p-1}\tau^{\delta^{-1}(\gamma s^2-2rs)}|e_r\rangle\langle
e_s|,\nonumber\\
V_F&=&V_{F_1}V_{F_2}=\sum\limits _{s=0}^{p-1}\tau^{\alpha\gamma
s^2}|e_{\alpha s}\rangle\langle e_s|.
\end{eqnarray}
If $\mathrm{det}(F)=-1$, then $\mathrm{det}(JF)=1$, and
$(JF,\chi)\in \mathrm{SL}(2,p)\ltimes (Z_p)^2$. Hence the
isomorphism images of elements in $\mathrm{ESL}(2,p)\ltimes(Z_p)^2$
can be determined once the images of elements in
$\mathrm{SL}(2,p)\ltimes(Z_p)^2$  and that of  $(J,\bm{0})$ are
determined respectively. The isomorphism image of $(J,\bm{0})$ is
the complex conjugation operator $\hat{J}$ \cite{App05}.

Following Appleby, $[F,\chi]$ is used to denote the isomorphism
image of $(F,\chi)$ under the correspondence \eref{eq:isomorphism1}
throughout this  paper. In the following discussion, we focus on the
Clifford group, except when otherwise stated.

Since  the Clifford group and $\mathrm{SL}(2,p)\ltimes (Z_p)^2$ are
isomorphic, they have the same class structure. There are also
$2p+4$ classes in the Clifford group, and the class representatives
can be chosen as the isomorphism images of that of
$\mathrm{SL}(2,p)\ltimes (Z_p)^2$ listed in \tref{tab:class2}. There
are $p+2$ classes with elements of order $p$,  two classes with
elements of order $2p$, one class with elements of order 2, and
$\frac{1}{2}\varphi(k)$ classes with elements of order $k$ for each
divisor of $p-1$ or
 $p+1$ which is not equal to 2.  In addition, if $p>3$, all order 3 elements are conjugated to
 each other, recovering the result obtained by Flammia \cite{Fla06}, so are all order 4 elements and order 6
 elements. By contrast, there are more than one classes if  the order is a divisor of $p-1$ or $p+1$ which is equal to 5 or
 larger than 6.

The spectrum, and in particular the dimension of each eigenspace of
a Clifford unitary plays an important role in proving our main
results in the next section.  Here we give a brief account of the
spectrum of the elements in each conjugacy class, see \cite{App09}
for additional information.

The spectrum  of each element in the class
$\Bigl[1,\Bigl(\begin{array}{c}
                     1\\
                    0
                  \end{array}
 \Bigr)\Big]$
is the same as that of $Z$, and  is thus nondegenerate. For each
element in the class
 $[z,\bm{0}]$,  according to \eref{eq:VF2}, there are two distinct eigenvalues $\pm1$ (all eigenvalues are defined up to an overall phase factor) with
 multiplicity $\frac{p\pm1}{2}$ respectively. For each element in the
 class $[a,\bm{0}]$, all eigenvalues are $(p-1)$st roots of
 unity; the eigenvalue 1 is doubly degenerate, and all other eigenvalues
 are nondegenerate. The spectrum of each element in the class $[a^l,
 \bm{0}]$ for $1\leq l\leq \frac{p-3}{2}$ is simply the corresponding power of that of each element in the class
 $[a,\bm{0}]$.

The spectrum of each element in the class
 $[b,\bm{0}]$ can be determined in virtue of  representation theory.
 Note that each  element
 $F=\Bigl(
     \begin{array}{cc}
       \alpha & \beta \\
       \gamma & \delta \\
     \end{array}
   \Bigr)$
in the class $b$ of $\mathrm{SL}(2,p)$ satisfies the two
inequalities, $\beta\neq0$ and $\alpha+\delta\neq2$. According to
\eref{eq:VF1},
\begin{eqnarray}
|\mathrm{tr}(V_F)|^2=\frac{1}{p}\Biggl|\sum_{s=0}^{p-1}\tau^{\beta^{-1}(\alpha+\delta-2)s^2}\Biggr|^2=1,
\end{eqnarray}
 similarly,
$\bigl|\mathrm{tr}\bigl[(V_F)^n\bigr]\bigr|^2=|\mathrm{tr}(V_{F^n})|^2=1$
for $n=1,\ldots,p$. If we take $(V_F)^n$ for $n=0,\ldots, p$ as a
representation of a cyclic group of order $p+1$, then the sum of
squared multiplicities of all irreducible components in this
representation is given by \cite{JL01}
\begin{eqnarray}
\label{eq:character1}
\frac{1}{p+1}\sum_{n=0}^p\bigl|\mathrm{tr}\bigl[(V_F)^n\bigr]\bigr|^2=\frac{p^2+p}{p+1}=p.
\end{eqnarray}
Since all irreducible representations of any cyclic group are one
dimensional,  the above representation  is a direct sum of $p$
one-dimensional irreducible representations. \Eref{eq:character1}
implies that  all the $p$ irreducible components are  distinct,
which in turn implies the nondegeneracy of the spectrum of $V_F$.
Each eigenvalue of $V_F$ is a $(p+1)$st root of unity, hence the
spectrum of $V_F$ contains all but one $(p+1)$st roots of unity. The
spectrum of each element in the class $[b^m,\bm{0}]$ for $1\leq
m\leq \frac{p-1}{2}$ is the corresponding power of the spectrum of
each element in the class $[b, \bm{0}]$.

Now consider each element in either of the class $[c_1,\bm{0}]$ or
$[c_2,\bm{0}]$. According to \eref{eq:VF2}, if
$\beta=0,\alpha=\delta=1$, that is
 $F\in Q_1$ ($Q_1$  is a Sylow $p$-subgroup of $\mathrm{SL}(2,p)$, see \eref{eq:Q1} for its definition),
  then $V_F$ is diagonal with diagonal entries $\tau^{\gamma
 s^2}$ for $s=0,\ldots,(p-1)$. The distinct eigenvalues of
 $V_F$ are $\tau^{\gamma
 s^2}$ for $s=0,\ldots,\frac{p-1}{2}$, and  all eigenvalues are doubly degenerate except the
 eigenvalue 1. Thus  for each  element in the class $[c_1,\bm{0}]$, the distinct eigenvalues are
 $\tau^{ s^2}$ for $s=0,\ldots,\frac{p-1}{2}$, and for each  element in the class $[c_2,\bm{0}]$, they are $\tau^{\nu
 s^2}$ for $s=0,\ldots,\frac{p-1}{2}$ ,
  where $\nu$ is any element in $Z_p^*$ that is not a quadratic
 residue. For each element in either of the two classes,
all eigenvalues  are doubly degenerate except
 the eigenvalue 1.

 For each element in either of the class
 $\Bigl[c_1,\Bigl(\begin{array}{c}
                     k_1\\
                    0
                  \end{array}
 \Bigr)\Bigr]$
 or
  $\Bigl[c_2,\Bigl(\begin{array}{c}
                     k_2\\
                    0
                  \end{array}
 \Bigr)\Bigr]$,
 direct inspection shows that its spectrum is the same as that of
 $Z$.

 For each element in either the class  $[zc_1,\bm{0}]$ or $[zc_2,\bm{0}]$,  according to
 \eref{eq:VF2}, all eigenvalues are distinct,  which are given by
 1, $ \pm\tau^{s^2}$, or 1, $\pm\tau^{\nu
 s^2}$, for $s=0,\ldots,\frac{p-1}{2}$.
The nondegeneracy of eigenvalues can also be shown
  using  similar arguments as
 applied to the elements in the class  $[b,\bm{0}]$. Suppose $F=zc_1$ or $F=zc_2$,
according to \eref{eq:VF2},
 $\bigl|\mathrm{tr}\bigl[(V_F)^{2k-1}\bigr]\bigr|^2=1$ for
 $k=1,\ldots,p$,
 and $\bigl|\mathrm{tr}\bigl[(V_F)^{2k}\bigr]\bigr|^2=p$ for
 $k=1,\ldots,(p-1)$.
 If we take $(V_F)^{n}$ for $n=0,\ldots,(2p-1)$ as a representation of a cyclic group of order
 $2p$, then the sum of  squared multiplicities of all irreducible components in
 this representation is given by
 \begin{eqnarray}
\frac{1}{2p}\sum_{n=0}^{2p-1}\bigl|\mathrm{tr}\bigl[(V_F)^{n}\bigr]\bigr|^2=\frac{p^2+p+p(p-1)}{2p}=p.
 \end{eqnarray}
Hence, each irreducible component occurs only once, which implies
the nondegeneracy of the spectrum of  $V_F$.

A unitary matrix is a \emph{monomial matrix} or \emph{in monomial
form} if there is only one nonzero entry in each row and each
column. A unitary group is a \emph{monomial group} if there exists a
unitary transformation that simultaneously turns all elements in the
group into monomial form. If every element in a monomial group is
already monomial, then the group is \emph{in monomial form}.
According to \eref{eq:VF2}, $V_F$ is monomial if $\beta=0$, that is
$F\in N_1$ ($N_j$s are defined in \sref{sec:SL}),  and it is a
permutation matrix if in addition $\gamma=0$.

Let $H$ be a subgroup of the Clifford group that contains the HW
group $D$.  The quotient group $H/D$ can be identified with a
subgroup of $\mathrm{SL}(2,p)$. If $H/D\subset N_j$  with $1\leq
j\leq p+1$, then $H$ is monomial, and it is already in monomial form
if in addition $j=1$. What is not so obvious is that if $H$ is
monomial, then $H/D$ is a subgroup of $N_j$ with $1\leq j\leq p+1$.
To see this, let $U$ be a unitary operator that brings $H$ into the
monomial form $H^\prime$, and $D^\prime$ be the image of $D$ under
the same transformation. Since $D^\prime$ can be turned into  $D$
with a suitable monomial unitary transformation, without loss of
generality, we can assume that $D^\prime=D$ and $U$  is a Clifford
unitary. Hence, $H^\prime/D\in N_1$, and  $H\in N_j$ with $1\leq
j\leq p+1$. As a consequence, when $H$ is monomial, $H/D$ is cyclic
if its order is equal to $2p$ or not a multiple of $p$, according to
the discussion in \sref{sec:SL}.

Zauner's conjecture states that HW fiducial vectors exist in any
finite dimension, and every such fiducial vector is an eigenvector
of a canonical order 3 unitary  \cite{Zau99, RBSC04, App05, SG10}
(there are several different versions of Zauner's conjecture
\cite{App05},  a specific one has been chosen here). Interestingly,
when $3|(p-2)$, Zauner's conjecture implies that the symmetry group
$G_{\mathrm{sym}}$ of a SIC~POVM cannot be monomial. To demonstrate
this point, let $G^\prime_{\mathrm{sym}}$ be the intersection of
$G_{\mathrm{sym}}$ with the Clifford group. If each fiducial vector
is stabilized by an order 3 element in the Clifford group, then  3
divides $|G^\prime_{\mathrm{sym}}/D|$, which in turn divides $|N_1|$
if $G^\prime_{\mathrm{sym}}$ is monomial, according to the previous
discussions. This, however, contradicts the fact that $|N_1|=p(p-1)$
is not divisible by 3.

\subsection{\label{sec:HWs}Heisenberg--Weyl groups in the Clifford group}
There are many other subgroups in the Clifford group that are
unitarily equivalent to the HW group defined in \eref{HW}; these
groups will also be called HW groups.  The normalizer of each of
these  groups will be  referred to as the Clifford group of that HW
group. If necessary, we will refer to the HW group defined in
\eref{HW} as the \emph{standard HW group}, and its (extended)
Clifford group as the \emph{standard (extended) Clifford group}. In
this section, we  focus on  these additional HW groups and Clifford
groups, since they play an important role in understanding the
structure of SIC~POVMs, as we shall see in \sref{sec:main} and
\sref{sec:sicpom3}.

For prime dimensions, each HW groups is  a $p$-group, and is thus
 contained in a Sylow $p$-subgroup \cite{KS04} of the Clifford group. In
correspondence to the $p+1$ Sylow $p$-subgroups $Q_j$  in
$\mathrm{SL}(2, p)$, there are $p+1$ Sylow $p$-subgroups $\bar{P}_j
(P_j)$ for $j=1,\ldots, p+1$ in the Clifford group
$\bar{\mathrm{C}}(p)$ ($\mathrm{C}(p)$), such that
$\bar{P}_j/\bar{D}=Q_j$ ($P_j/D=Q_j$). The intersection of these
Sylow $p$-subgroups is exactly the standard HW group.

Since all Sylow $p$-subgroups in the Clifford group are conjugated
to each other, it suffices to study any one of them, say the Sylow
$p$-subgroup $\bar{P}_1$ ($P_1$), which is generated by the
following three elements:
\begin{eqnarray}
\fl V&=&\left[\left(
      \begin{array}{cc}
        1 & 0 \\
        1 & 1 \\
      \end{array}
    \right),\left(
              \begin{array}{c}
                0 \\
                0 \\
              \end{array}
            \right)
    \right], \qquad
   X= \left[\left(
      \begin{array}{cc}
        1 & 0 \\
        0 & 1 \\
      \end{array}
    \right),\left(
              \begin{array}{c}
                1 \\
                0 \\
              \end{array}
            \right)
    \right],\qquad
    Z=\left[\left(
      \begin{array}{cc}
        1 & 0 \\
        0 & 1 \\
      \end{array}
    \right),\left(
              \begin{array}{c}
                0 \\
                1 \\
              \end{array}
            \right)
    \right].
\end{eqnarray}
The order of $\bar{P}_1$ ($P_1$) is $p^3$ ($p^4$),   and the order
of any element in $\bar{P}_1$ other than identity is $p$. The centre
of $\bar{P}_1$ is the cyclic group $\langle Z\rangle$ generated by
$Z$, while the centre of $P_1$ is the cyclic group $\langle\omega
I\rangle$ generated by $\omega I$. Since each subgroup of $P_1$ of
order $p^3$ necessarily contains the subgroup $\langle\omega
I\rangle$, there is a one-to-one correspondence between the
subgroups of $P_1$ of order $p^3$ and subgroups of $\bar{P}_1$ of
order $p^2$. There are $p+1$ order $p^2$ subgroups in $\bar{P}_1$,
$\langle Z, V^j X\rangle$ for $j=0, \ldots, p-1 $ and $\langle Z,
V\rangle$. The first $p$ of them are unitarily equivalent, as we
shall see shortly.

According to \eref{eq:VF2},
\begin{eqnarray}
     V=\mathrm{diag}\bigl(1,\tau, \tau^4,\ldots,
     \tau^{(p-1)^2}\bigr),\qquad
\mathrm{det}(V)=\tau^{p(p-1)(2p-1)/6}.
\end{eqnarray}
If $p\geq5$, $(p-1)(2p-1)$ is divisible by $6$, and
$\mathrm{det}(V)$ is equal to $\tau^{p(p-1)(2p-1)/6}=1$, recall that
$\tau=-\mathrm{e}^{\pi \rmi/p}$. Define
\begin{eqnarray}\label{eq:HWpermute1}
U&=&\mathrm{diag}(1,\mathrm{e}^{\rmi\phi_1},\ldots,\mathrm{e}^{\rmi\phi_{p-1}}),\nonumber\\
\mathrm{e}^{\rmi\phi_{1}}&=&\tau, \nonumber\\
\mathrm{e}^{\rmi\phi_{2}}&=&\tau^{1+2^2},\nonumber\\
&\ldots&\nonumber\\
\mathrm{e}^{\rmi\phi_{p-1}}&=&\tau^{1+2^2+\cdots+(p-1)^2}=1;
\end{eqnarray}
then we have
\begin{eqnarray}
U^jZU^{j\dag}=Z,\qquad  U^jXU^{j\dag}=V^j{X},\qquad \mbox{for}\quad
j=0,\ldots,p-1.
\end{eqnarray}
If $p=3$, then
$\mathrm{det}(V)=\tau^{p(p-1)(2p-1)/6}=\tau^5=\mathrm{e}^{2\pi
\rmi/3}$. Define $V^\prime=\mathrm{e}^{4\rmi\pi/9}V$ and
\begin{eqnarray}
\label{eq:HWpermute2}
U&=&\mathrm{diag}(1,\mathrm{e}^{-2\rmi\pi/9},\mathrm{e}^{-4\rmi\pi/9});
\end{eqnarray}
then we have  $\mathrm{det}(V^\prime)=1$ and
\begin{eqnarray}\label{eq:HWpermute3}
&&U^jZU^{j\dag}=Z,\qquad U^jXU^{j\dag}=V^{\prime j}{X},\qquad
\mbox{for}\quad j=0, 1, 2.
\end{eqnarray}

In conclusion, all the $p$ groups  $\langle Z, V^j X\rangle$ for
$j=0, \ldots, p-1 $ are unitarily equivalent to the standard HW
group, and they are permuted cyclically under the transformation $U$
in \eref{eq:HWpermute1} for $p\geq5$ or in \eref{eq:HWpermute2} for
$p=3$. The group $\langle Z, V\rangle$ cannot be unitarily
equivalent to the HW group because all elements in the group are
diagonal. Hence, there are  $p(p+1)+1$ order $p^2$ ($p^3$) subgroups
in the Clifford group $\bar{\mathrm{C}}(p)$ ($\mathrm{C}(p)$), out
of which $p^2$ subgroups are unitarily equivalent to the HW group,
recall that the standard HW group is the intersection of the  $p+1$
Sylow $p$-subgroups of the Clifford group.

The $p^2-1$ additional HW groups in the Clifford group form a single
orbit if $p=3$ or $3|(p-2)$, and three orbits if $3|(p-1)$. To
demonstrate this point, it is enough to show that the $p-1$
additional HW groups in the Sylow $p$-subgroup $\bar{P}_1$ form the
corresponding number of orbits in the two cases respectively,
because all Sylow $p$-subgroups are conjugated to each other.
Suppose the two  HW groups $\langle Z, V^l X\rangle$ and $\langle Z,
V^jX\rangle$, where $1\leq l, j\leq p-1$, are connected by the
Clifford unitary $\Bigl[\Bigl(
    \begin{array}{cc}
      \alpha & \beta \\
      \gamma & \delta \\
    \end{array}
  \Bigr),\Bigl(
  \begin{array}{c}
    k_1 \\
    k_2 \\
  \end{array}
\Bigr)\Bigr]$; then the Clifford unitary  belongs to the normalizer
of $\bar{P}_1$, and $\Bigl(
    \begin{array}{cc}
      \alpha & \beta \\
      \gamma & \delta \\
    \end{array}
  \Bigr)$
belongs to the normalizer $N_1$ of $Q_1$ (see \eref{eq:Q1} and
\eref{eq:N1} for the definitions of $Q_1$ and $N_1$ respectively),
which implies that  $\beta=0$, and $\delta=\alpha^{-1}$. Recall that
$V^l X=\Bigl[\Bigl(
           \begin{array}{cc}
             1 & 0 \\
             l & 1 \\
           \end{array}
         \Bigr)
,\Bigl(
            \begin{array}{c}
             1  \\
              l\\
            \end{array}
          \Bigr)
\Bigr]$, we have
\begin{eqnarray}\label{eq:HWnormalizer}
\fl V_1&=& \left[\left(
      \begin{array}{cc}
        \alpha & 0 \\
        \gamma & \alpha^{-1} \\
      \end{array}
    \right),\left(
            \begin{array}{c}
              k_1 \\
              k_2 \\
            \end{array}
          \right)
\right]V^l X\left[\left(
      \begin{array}{cc}
        \alpha & 0 \\
        \gamma & \alpha^{-1} \\
      \end{array}
    \right),\left(
            \begin{array}{c}
              k_1 \\
              k_2 \\
            \end{array}
          \right)
\right]^{-1}=\left[\left(
           \begin{array}{cc}
             1 & 0 \\
             \alpha^{-2}l & 1 \\
           \end{array}
         \right)
,\left(
            \begin{array}{c}
              \alpha  \\
              l^\prime\\
            \end{array}
          \right)
\right],
\end{eqnarray}
where $l^\prime\in Z_p$, whose specific value is not important here.
Hence, $V_1\in \langle Z, V^jX\rangle$ if and only if
$\alpha^{-3}l=j$. If $p=3$ or $3|(p-2)$, then $\alpha^{-3}$ may take
any value in $Z_p^*$, and there exists $\alpha$ satisfying
$\alpha^{-3}l=j$ for any pair $l,j\in Z_p^*$. As a result, the
$(p-1)$ HW groups $\langle Z, V^lX\rangle$ for $l=1,\ldots,p-1$ are
on the same orbit. If $3|(p-1)$, then $\alpha^{-3}$ may only take
one third possible values in $Z_p^*$. Hence, the $(p-1)$ HW groups
form three orbits of equal length $\frac{p-1}{3}$.

Suppose $\bar{D}^\prime$ is any HW group in $\bar{P}_1$ other than
the standard one, and $\bar{\mathrm{C}}^\prime(p)$  its  Clifford
group. Our analysis in the last paragraph also shows that the
normalizer of $\bar{D}^\prime$ within $\bar{\mathrm{C}}(p)$ consists
of the following elements:
\begin{eqnarray}
\left[\left(
      \begin{array}{cc}
        \alpha & 0 \\
        \gamma & \alpha^{-1} \\
      \end{array}
    \right),\left(
            \begin{array}{c}
              k_1 \\
              k_2 \\
            \end{array}
          \right)
\right]\qquad \mbox{with}\quad \alpha,\gamma, k_1,k_2\in Z_p,\quad
\alpha^3=1.
\end{eqnarray}
If $p=3$ or $3|(p-2)$, this group is exactly $\bar{P}_1$, that is,
$\bar{\mathrm{C}}^\prime(p)\cap \bar{\mathrm{C}}(p)=\bar{P}_1$; if
$3|(p-1)$, $\bar{P}_1$ is a normal subgroup of
$\bar{\mathrm{C}}^\prime(p)\cap \bar{\mathrm{C}}(p)$ with index $3$.

We shall prove in the next section that the symmetry group
$G_{\mathrm{sym}}$ of a group covariant SIC~POVM is a subgroup of
some Clifford group for  any prime dimension.  It follows from the
above discussion that the number of HW groups in $G_{\mathrm{sym}}$
may only take three possible values $1,p, p^2$; that is, the
SIC~POVM may only be covariant with  respect to $1, p$ or $p^2$ HW
groups. If $|G_{\mathrm{sym}}|$ is not divisible by $p^4$, then
$G_{\mathrm{sym}}$ contains only one HW group. Otherwise, each Sylow
$p$-subgroup of $G_{\mathrm{sym}}$ is also a Sylow $p$-subgroup of
the Clifford group containing $G_{\mathrm{sym}}$. In addition,  the
number of Sylow $p$-subgroups in $G_{\mathrm{sym}}$ is either $1$ or
$p+1$, according to Sylow's theorem \cite{KS04}, so the number of HW
groups in $G_{\mathrm{sym}}$ is either $p$ or $p^2$.

\section{\label{sec:main}The symmetry group of any group covariant SIC~POVM in any prime dimension is a subgroup of certain Clifford group }

In this section, we prove  that the (extended) symmetry group of any
group covariant SIC~POVM in any prime dimension is a subgroup of
certain (extended) Clifford group, and  derive  simple criteria on
determining whether two group covariant SIC~POVMs are unitarily or
antiunitarily equivalent. First, we show that a group covariant
SIC~POVM in any prime dimension is covariant with respect to the HW
group. Then we prove our main result in three cases separately,
namely the special case $p=2$, the general case $p\geq5$, and
 the special case $p=3$.
\subsection{\label{sec:HWcovariance}A group covariant SIC~POVM in any prime dimension is covariant with respect to the Heisenberg--Weyl group}
In this section, we prove the following lemma.
\begin{lemma}\label{lem:HWcovariance}
In any prime dimension, a group covariant SIC~POVM is necessarily
covariant with respect to the HW group.
\end{lemma}

First, we need the following proposition.
\begin{proposition}\label{pro:nonabelian}
In any finite dimension, if a SIC~POVM is covariant with respect to
a unimodular unitary group $G$, then $G$ is necessarily nonabelian.
In particular, the symmetry group of any group covariant SIC~POVM is
necessarily nonabelian.
\end{proposition}
Suppose there exists a SIC~POVM in dimension $d$ which is covariant
with respect to an abelian unimodular unitary group $G$, which may
be assumed to be diagonal, without loss of generality. Let
$|\psi_j\rangle=(a_{j1}, \ldots, a_{jd})^T$ for $j=1, \ldots, d^2$
be the $d^2$ vectors in the SIC~POVM. Since $|\psi_j\rangle$s are
related to each other by the diagonal unitary group $G$, the modulus
of a given entry is independent of the vectors; that is, $|a_{jk}|$
is independent of $j$. Hence, the condition  $\sum_{j=1}^{d^2}
|\psi_j\rangle\langle\psi_j|=d I$ implies that $|a_{jk}|=1/\sqrt{d},
\forall j, k$. Suppose $U=\mathrm{diag}(u_1, \ldots, u_d)$ is an
element in $G$ which does not stabilize $|\psi_1\rangle$; then we
have, according to \eref{eq:SIC},
\begin{eqnarray}
d^2|\langle\psi_1|U|\psi_1\rangle|^2=\Biggl|\sum_{j=1}^{d}u_j\Biggl|^2=\frac{d^2}{d+1}.
\end{eqnarray}
Since $u_j$s are  roots of unity, which are algebraic integers, the
expression in the middle of the above equation is also an algebraic
integer. On the other hand, $d^2/(d+1)$ cannot be an algebraic
integer since it is a fraction which is not an integer \cite{JL01},
a contradiction. This completes the proof of the  first part of
proposition~\ref{pro:nonabelian}. The second part of
proposition~\ref{pro:nonabelian}  follows immediately, since  a
SIC~POVM is group covariant if and only if it is covariant with
respect to its symmetry group.

Let us now turn back to  the proof of lemma~\ref{lem:HWcovariance}.
In any prime dimension, a group covariant SIC~POVM is also
necessarily covariant with respect to any Sylow $p$-subgroup, say
$P$, of its symmetry group $G_{\mathrm{sym}}$. Since $P$ must be
nonabelian according to proposition~\ref{pro:nonabelian},
 the order of $P$ is at least $p^3$; recall that all groups of
order $p$ or $p^2$ are abelian. In addition, $P$ must be irreducible
when taken as a representation of itself, because the degree of any
irreducible representation of a finite group divides its order
\cite{JL01}. The centre of $P$ has order at least $p$ because a
$p$-group has a nontrivial centre \cite{KS04}. Since $P$ is
irreducible, any element in its centre  is proportional to the
identity, which, together with the unimodular condition, implies
that the centre is the cyclic group $\langle \omega I\rangle$
generated by $\omega I$. Since the $p$-group $P/\langle\omega
I\rangle$ also  has a nontrivial centre, there exists a nontrivial
element $X^\prime$ in $P$ such that $X^\prime\langle\omega I\rangle$
commutes with all elements in $P/\langle\omega I\rangle$. Hence,
there exists another element $Z^\prime$ in $P$, such that $ Z^\prime
X^\prime Z^{\prime-1}X^{\prime-1}=\omega^k I$ with $1\leq k<p$. In
addition, $k$ can be chosen to be 1.

Since any irreducible representation of a $p$-group is monomial
\cite{CR62}, we can assume that $P$ is in monomial form,  without
loss of generality. Note that at least one of the two elements, say
$Z^\prime$, is not diagonal, and that each element of $P$ which is
not diagonal necessarily has the same spectrum as that of $Z$. We
can choose a new basis such that $Z^\prime=\mathrm{diag}(1, \omega,
\ldots, \omega^{p-1})$, while leaving the commutation relation
between $Z^\prime$ and $X^\prime$ unchanged. It follows that
$X^\prime X^{-1}$ commutes with $Z^\prime$, and is thus diagonal. We
can turn $X^\prime$ into $X$ with a suitable diagonal unitary
transformation which leaves $Z^\prime$ unchanged. Hence,  the
subgroup of $P$ generated by the two elements $X^\prime$ and
$Z^\prime$ is unitarily equivalent to the HW group defined in
\eref{HW}. We may now identify $X^\prime$ and $Z^\prime$ with $X$
and $Z$ respectively, and call the  group generated by the two
elements  HW group. It remains to show that the SIC~POVM is
covariant with respect to this HW group, which is guaranteed if, for
each vector in the SIC~POVM, the stability group within this HW
group does not contain any nontrivial element. Suppose otherwise,
without loss of generality, we can assume that a vector
$|\psi\rangle$ of the SIC~POVM is stabilized by $Z$. Then
$|\psi\rangle$   can only have one nonzero entry, which implies that
$X|\psi\rangle$ and $|\psi\rangle$ are orthogonal to each other, a
contradiction. This completes the proof of
lemma~\ref{lem:HWcovariance}. Hence, for any SIC~POVM in any prime
dimension, group covariance is equivalent to HW covariance.

\subsection{\label{sec:d2}Special case p=2}

In dimension two,  the Clifford group $\bar{\mathrm{C}}(2)$ is
generated by the Hadamard operator $\frac{1}{\sqrt{2}}\Bigl(
  \begin{array}{cc}
    1 & 1 \\
    1 & -1 \\
  \end{array}
\Bigr)$ and phase operator $\Bigl(
  \begin{array}{cc}
    1 & 0 \\
    0 & \rmi \\
  \end{array}
\Bigr)$ \cite{NC00}, and the extended Clifford group is generated by
the complex conjugation operator $\hat{J}$ in addition to the two
operators. The orders of the Clifford group and the extended
Clifford group are $24$ and $48$ respectively. There is only one
orbit of fiducial vectors under either the Clifford group or the
extended Clifford group. One of the fiducial vectors is given by
\begin{eqnarray}
|\psi\rangle=\left(
               \begin{array}{c}
                 \sqrt{\frac{3+\sqrt{3}}{6}} \\
                 \rme^{\rmi\pi/4}\sqrt{\frac{3-\sqrt{3}}{6}} \\
               \end{array}
             \right).
\end{eqnarray}
The order of the stability group of each fiducial vector within the
Clifford group (extended Clifford group) is 3 (6); thus there are
eight fiducial vectors constituting two SIC~POVMs \cite{Zau99,
RBSC04, App05}.

When represented  on the Bloch sphere, the eight fiducial vectors
constitute a cube, and the two SIC~POVMs constitute two regular
tetrahedra respectively, which are related to each other by space
inversion. The Clifford group corresponds to the rotational symmetry
group of the cube, while the extended Clifford group corresponds to
the full symmetry group of the cube. The extended symmetry group of
each SIC~POVM corresponds to the full symmetry group of the
tetrahedron, which is a subgroup of the full symmetry group of the
cube. Hence, the extended symmetry group of each SIC~POVM contains
only one HW group, and it is a subgroup of the extended Clifford
group.

Moreover,  all SIC~POVMs in dimension two are unitarily equivalent,
since any SIC~POVM, when represented on the Bloch sphere,
corresponds to a regular tetrahedron. Hence, any SIC~POVM in
dimension two is covariant with respect to a unique HW group, and
its (extended) symmetry group is a subgroup of the (extended)
Clifford group.

\subsection{\label{sec:general}General case $p\geq5$}
For  $p\geq5$, the following theorem proved by Sibley \cite{Sib74}
is crucial to our later discussion.
\begin{theorem}\label{thm:Sibley}
$\mathrm{(Sibley)}$ Suppose G is a finite group with a faithful,
irreducible, unimodular and quasiprimitive representation of prime
degree $p\geq5$. If a Sylow p-subgroup $P$ of G has order $p^3$,
then $P$ is normal in G, and G/P is isomorphic to a subgroup of
$\mathrm{SL}(2,p)$.
\end{theorem}
A quasiprimitive representation is one whose restriction to every
subgroup is homogeneous, that is a multiple of one irreducible
representation of the subgroup. An irreducible representation of
prime degree that is not quasiprimitive is monomial \cite{Sib74}.

Let $G_{\mathrm{sym}}$ be the symmetry group of a group covariant
SIC~POVM in any prime dimension $p\geq5$. Taken as a representation
of itself, $G_{\mathrm{sym}}$ is irreducible, because it contains
the HW group $D$, which is irreducible.  To apply  Sibley's theorem,
we shall first prove that the order of each Sylow $p$-subgroup of
$G_{\mathrm{sym}}$ is $p^3$.
 Suppose otherwise, then the HW group $D$ is a proper subgroup
of one of the Sylow $p$-subgroups. The normalizer $N(D)$ of $D$ in
this Sylow $p$-subgroup is strictly larger than $D$, and $N(D)/D$
contains a subgroup of order $p$ \cite{KS04}. It follows that $N(D)$
contains a subgroup of order $p^4$ which in turn contains $D$ as a
normal subgroup. This group of order $p^4$ is also a Sylow
$p$-subgroup of the Clifford group, and can be taken  as $P_1$
without loss of generality, since all Sylow $p$-subgroups are
conjugated to each other. Suppose $V\in\Bigl[\Bigl(
\begin{array}{cc}
  1 & 0 \\
  1 & 1
\end{array}\Bigr),\bm{0}\Bigr]
$ is  unimodular; then  the group $H=\langle \omega I, Z, V\rangle $
is a normal subgroup of $P_1$ consisting of  its diagonal elements.
Since the order of the stability group within $H$ of each fiducial
vector is the same, and the  action cannot be transitive according
to proposition~\ref{pro:nonabelian}, the fiducial vectors form $p$
orbits of equal length $p$. It follows that each fiducial vector,
say $|\psi\rangle$, is stabilized by some nontrivial element of the
form $V^j Z^k$, with $1\leq j\leq p-1$ and $0\leq k\leq p-1$. Since
$V^j Z^k$ is diagonal,   all $p$ fiducial vectors $Z^l|\psi\rangle$
for $l=0,\ldots,p-1$ are simultaneously stabilized by it. These
fiducial vectors must belong to a same eigenspace of
 $V^j Z^k$. However,
 each eigenspace of $V^j Z^k$ for $1\leq j\leq p-1$ has dimension  of at most two as shown
in \sref{sec:Clifford}, and thus cannot admit more than two fiducial
vectors when $p\geq5$; hence  a contradiction would arise. In
conclusion, the order of each Sylow $p$-subgroup of
$G_{\mathrm{sym}}$ is $p^3$, and
 $D$ is a Sylow $p$-subgroup of $G_{\mathrm{sym}}$.

If $G_{\mathrm{sym}}$ is  quasiprimitive, then   $D$ is a normal
subgroup of $G_{\mathrm{sym}}$, or equivalently, $G_{\mathrm{sym}}$
is a subgroup of the Clifford group of $D$, according to Sibley's
theorem. In addition, $G_{\mathrm{sym}}$ contains only one HW group.

Now suppose $G_{\mathrm{sym}}$ is in monomial form. By a suitable
monomial unitary transformation and a different choice of generators
if necessary, we can always keep the HW group in the standard form.
Let $T$ be the normal subgroup of $G_{\mathrm{sym}}$ which consists
of its diagonal elements, and $S=G_{\mathrm{sym}}/T$. Note that
$|S|$ is divisible by $p$, otherwise the SIC~POVM would be covariant
with respect to the abelian group $T$, which contradicts
proposition~\ref{pro:nonabelian}. Hence, $|T|$ is not divisible by
$p^3$, since $|G_{\mathrm{sym}}|$  is not  divisible by $p^4$ as
shown previously. In addition, under the action of $T$, the fiducial
vectors of the SIC~POVM form $p$ orbits of equal length $p$, and two
fiducial vectors are on the same orbit generated by $T$ if and only
if they are on the same orbit generated by $\langle Z\rangle$.

To show that $D$ is a normal subgroup of $G_{\mathrm{sym}}$, we
shall first show that $T$ contains no other elements except those
generated by $Z$ and $\omega I$; that is, $T=\langle \omega I,
Z\rangle$ with $|T|=p^2$. Suppose $|T|>p^2$, then $T$ contains a
nontrivial element $U$ whose order is not a multiple of $p$. Since
$U$ cannot stabilize all fiducial vectors,  there exists at least
one fiducial vector, say $|\psi\rangle$, not stabilized by $U$. On
the other hand, since $U|\psi\rangle$ and $|\psi\rangle$ are on the
same orbit of $T$ and hence on the same orbit of $\langle Z\rangle$,
we have $U|\psi\rangle=\rme^{\rmi\phi}Z^k|\psi\rangle$, where $1\leq
k\leq p-1$, and $\rme^{\rmi\phi}$ is an overall phase factor. Note
that $|\psi\rangle$
 has at least two nonzero entries; let $\rme^{\rmi\phi_1},
\rme^{\rmi\phi_2}$ be the two diagonal entries of $U$ corresponding
to any two nonzero entries of $|\psi\rangle$ respectively; then
$\rme^{\rmi(\phi_1-\phi_2)}$ is a primitive $p$th root of unity,
contradicting  the fact that the order of $U$ is not a multiple of
$p$.

Now we are ready to show that $D$ is a normal subgroup of
$G_{\mathrm{sym}}$. Let $U$ be an arbitrary element in
$G_{\mathrm{sym}}$.  Since $T=\langle\omega I, Z\rangle$ is a normal
subgroup of $G_{\mathrm{sym}}$, $UZU^{\dag}=\omega^{k_1}Z^{k_2}$ for
some integers $k_1, k_2$, it remains to show that $UXU^{\dag}\in D$,
or equivalently $U^{\dag}XU\in D$. According to the following
equalities:
\begin{eqnarray}
\label{eq:commute}
 &&U^{\dag}X U
Z(U^{\dag}XU)^{\dag}=U^{\dag}X\omega^{k_1}Z^{k_2}X^{\dag}U
=U^{\dag}\omega^{k_1-k_2}Z^{k_2}U=\omega^{-k_2}Z,
\end{eqnarray}
  $X^{-k_2}U^{\dag}X U $  commutes with $Z$, and hence belongs to
$T$, which in turn implies that
 $X^{-k_2}U^{\dag}X U=\omega^{k_1^\prime}Z^{k_2^\prime}$ for some integers $k_1^\prime, k_2^\prime$,
and  $U^{\dag}X U=\omega^{k_1^\prime}X^{k_2}Z^{k_2^\prime}\in D$.
Hence, when $G_{\mathrm{sym}}$ is monomial, $G_{\mathrm{sym}}$ is
also a subgroup of the Clifford group, and contains only one  HW
group. Moreover, in this case, the stability group of each fiducial
vector is cyclic, and there exists a fiducial vector on the same
orbit of the Clifford group whose stability group is generated by a
permutation matrix. To demonstrate this point, note that the
stability group of each fiducial vector is isomorphic to
$G_{\mathrm{sym}}/D$, and its order is not a multiple of $p$. Since
$G_{\mathrm{sym}}$ is monomial,
 $G_{\mathrm{sym}}/D$ is isomorphic to a subgroup of $N_1$ according
 to the discussions in
 \sref{sec:Clifford}.
 It follows that the stability group of each fiducial vector is
 cyclic; recall that all subgroups of $N_1$ whose order is not a multiple of $p$ are
 cyclic. In addition, the generator of the stability group
is conjugated to the class representative $[a^k,\bm{0}]$ with $0\leq
k\leq \frac{p-1}{2}$ (the isomorphism image of  the class
representative $(a^k, \bm{0})$ listed in \tref{tab:class2}).
According to \sref{sec:Clifford}, $[a^k,\bm{0}]$ is a permutation
matrix up to an overall phase factor, so there exists a fiducial
vector whose stability group is generated by a permutation matrix.
This observation may help search for SIC~POVMs with certain specific
symmetry.

In conclusion, for any prime dimension larger than three, each group
covariant SIC~POVM is covariant under one and only one HW group, and
its symmetry group is a subgroup of the Clifford group. The
conclusion can also be extended to cover antiunitary operations.
Note that any antiunitary operation in the extended symmetry group
of the SIC~POVM must preserve the HW group, and thus must belong to
the extended Clifford group. Also, the same results  hold for
dimension two according to the discussion in \sref{sec:d2}. So we
obtain
\begin{theorem}
\label{thm:main1} In any prime dimension not equal to three, each
group covariant SIC~POVM is covariant with respect to a unique HW
group. Furthermore, its (extended) symmetry group
 is a subgroup of the (extended) Clifford group.
\end{theorem}

To determine whether two given SIC~POVMs are equivalent can be a
very challenging task because there are infinitely many unitary
(antiunitary) operations to test. Usually, we need to find the
explicit transformation to claim that they are equivalent, and we
need to find some invariant that can distinguish the two SIC~POVMs
to claim that they are inequivalent. These two approaches will be
illustrated with SIC~POVMs in dimension three in \sref{sec:sicpom3}.
However, neither approach is easy in general. The difficulty is
reflected in the following long-standing open question:  Are two
SIC~POVMs on two different orbits of the (extended) Clifford group
equivalent? Fortunately, we can solve this open question for prime
dimensions not equal to three in virtue of theorem~\ref{thm:main1}.

Consider two HW covariant SIC~POVMs in any prime dimension not equal
to three. If there exists a unitary (antiunitary) operation which
maps one SIC~POVM to the other, it must preserve the HW group, and
hence belongs to the Clifford group (extended Clifford group). It
follows that the two SIC~POVMs are on the same orbit of the Clifford
group (extended Clifford group). So we obtain
\begin{corollary}\label{cor:generalEqui}
In any prime dimension not equal to three, two  SIC~POVMs covariant
with respect to the same HW  group are unitarily (unitarily or
antiunitarily) equivalent  if and only if they are on the same orbit
of the Clifford group (extended Clifford group).
\end{corollary}

In any prime dimension not equal to three, as an immediate
consequence of theorem~\ref{thm:main1} and
corollary~\ref{cor:generalEqui}, the different orbits of SIC~POVMs
found by Scott and Grassl \cite{SG10} are not unitarily or
antiunitarily equivalent. In particular, two HW covariant SIC~POVMs
cannot be  unitarily or antiunitarily equivalent if their respective
fiducial vectors have non-isomorphic stability groups (within the
extended Clifford group). For example, the two orbits of SIC~POVMs
in dimension seven discovered by Appleby \cite{App05} are not
unitarily or antiunitarily equivalent. However, it should be
emphasized that, without theorem~\ref{thm:main1} and
corollary~\ref{cor:generalEqui}, this seemingly obvious criterion is
not well justified {\it a priori}. We shall see counterexamples in
dimension three in \sref{sec:infinite}.

\subsection{\label{sec:special}Special case $p=3$}
Now consider the special case $p=3$. First, assume that
$G_{\mathrm{sym}}$ is not monomial. According to the classification
of finite linear groups of degree 3 by Blichfeldt \cite{Bli917}, the
order of the Sylow $p$-subgroup of $G_{\mathrm{sym}}$ is at most
$p^4$, and if it is equal to $p^4$, then $G_{\mathrm{sym}}$ is
isomorphic to some subgroup of the Clifford group. If the order of
the Sylow $p$-subgroup is $p^3$, there is a counterexample to
Sibley's theorem (theorem~\ref{thm:Sibley}). This counterexample is
a unimodular unitary group of order 1080 whose collineation group
(of order 360) is isomorphic to the alternating group of six letters
\cite{Bli917}. However, this group cannot be the symmetry group of
any SIC~POVM. Suppose there exists a SIC~POVM with this group as its
symmetry group; let $U$ be an order 5 element in the group. Under
the action of the group generated by $U$, the nine fiducial vectors
form disjoint orbits of length either 1 or $5$. It follows that
there are four orbits of length 1, that is, four fiducial vectors
stabilized by $U$. These four fiducial vectors must belong to a same
eigenspace of $U$; otherwise at least two  of them would be
orthogonal to each other. On the other hand, the dimension of this
eigenspace is at most two, because $U$ is not proportional to  the
identity. However, a two-dimensional subspace cannot admit four
fiducial vectors. Suppose otherwise, from the  Bloch sphere
representation, one can see that the maximum pairwise fidelity among
the four fiducial vectors is no smaller than $\frac{1}{3}$,
contradicting the fact that the pairwise fidelity among fiducial
vectors of a SIC~POVM in dimension three is equal to $\frac{1}{4}$.
In conclusion, $G_{\mathrm{sym}}$ must be a subgroup of some
Clifford group when it is not monomial.

Now suppose  that $G_{\mathrm{sym}}$ is in monomial form, and that
one of the HW groups contained in $G_{\mathrm{sym}}$ is in the
standard form, as in the case $p\geq5$. Let $T$ be the normal
subgroup of $G_{\mathrm{sym}}$ consisting of its diagonal elements,
and $S=G_{\mathrm{sym}}/T$.

If $T=\langle \omega I, Z\rangle$, then we can conclude that
$G_{\mathrm{sym}}$ contains only one HW group, and it is a subgroup
of the Clifford group, following a similar reasoning as that applied
to  the case  $p\geq 5$. However, it turns out that this situation
does not occur for the special case $p=3$ \cite{App05} (see also
\sref{sec:sicpom3}), in sharp contrast with the general case
$p\geq5$.

Otherwise, each fiducial vector is stabilized by some nontrivial
element in $T$. Let $|\psi\rangle$ be a fiducial vector, and $U$ a
nontrivial element in $T$ that stabilizes $|\psi\rangle$. Simple
analysis shows that two of the diagonal entries of $U$ must be
identical, and $|\psi\rangle$ must have two nonzero entries with
equal modulus $\frac{1}{\sqrt{2}}$ and a zero entry. With out loss
of generality, we may assume that
$U=\rme^{\rmi\phi^\prime}\mathrm{diag}(1,1,\rme^{\rmi\phi})$, and
$|\psi\rangle=\frac{1}{\sqrt{2}}(1,\rme^{\rmi t},0)$; indeed all
vectors of this form are fiducial vectors \cite{App05} (see also
\sref{sec:sicpom3}). To ensure that $UX|\psi\rangle$ be a fiducial
vector in the SIC~POVM, $\phi$ can only take two possible values
$\pm \frac{2\pi}{3}$. We can choose
$U=\rme^{-\rmi2\pi/9}\mathrm{diag}(1,1,\omega)$ for definiteness,
where $\phi^\prime$ has been chosen such that $U$ is unimodular. Now
it is straightforward to verify that $T$ cannot contain any elements
other than those generated by the following three elements, $\omega
I, Z, U$ and $|T|=27$.

The order of the  group $S$ may either be $3$ or $6$ and,
correspondingly, the order of $G_{\mathrm{sym}}$ is either $81$ or
$162$. If $|S|=3$, then $G_{\mathrm{sym}}$ is a $p$-group of order
$3^4$; hence the normalizer of $D$ in $G_{\mathrm{sym}}$ is strictly
larger than $D$ \cite{KS04}, which implies that $D$ is a normal
subgroup of $G_{\mathrm{sym}}$. If $|S|=6$, $G_{\mathrm{sym}}$
contains a Sylow $p$-subgroup $P$ of order $81$ and with index 2,
such that $D$ is a normal subgroup of $P$. Note that $P$ is also a
Sylow $p$-subgroup of the Clifford group, and thus contains two
other normal subgroups which are unitarily equivalent to $D$, or two
other HW groups, as shown in Sec~\ref{sec:HWs}. At least one of the
three HW groups is also  a normal subgroup of $G_{\mathrm{sym}}$. In
fact, according to our discussion in \sref{sec:HWs}, only one of the
three HW groups is normal in  $G_{\mathrm{sym}}$, and the other two
are conjugated to each other. Hence, when $G_{\mathrm{sym}}$ is
monomial, $G_{\mathrm{sym}}$ is also a subgroup of some Clifford
group. In addition, in this case, the stability group of each
fiducial vector is cyclic, because, according to
\sref{sec:Clifford}, $G_{\mathrm{sym}}/D$ is isomorphic to a
subgroup of $N_1$,  which is cyclic.

In conclusion, the symmetry group of any group covariant SIC~POVM in
dimension three is also a subgroup of some Clifford group. However,
it should be emphasized that a SIC~POVM may be covariant under more
than one HW groups. Furthermore, its  symmetry group may be a
subgroup of the Clifford group of one of  the HW groups but not a
subgroup of other Clifford groups, say the standard Clifford group.

According to \sref{sec:HWs}, the symmetry group
$\bar{G}_{\mathrm{sym}}$ of a SIC~POVM may contain three or nine  HW
groups if it contains more than one. The order of
$\bar{G}_{\mathrm{sym}}$ may only take three possible values 27, 54
or 216, and  cannot take the value 108, since the quotient group of
$\bar{G}_{\mathrm{sym}}$ with respect to its normal HW subgroup is
isomorphic to a subgroup of $\mathrm{SL}(2,3)$, which has no
subgroup of order 12 \cite{Mac75}. Let
$\bar{D}^{(1)},\ldots,\bar{D}^{(k)}$ be the HW groups contained in
$\bar{G}_{\mathrm{sym}}$, and $\bar{\mathrm{C}}^{(1)}(p),\ldots,
\bar{\mathrm{C}}^{(k)}(p)$ their associated Clifford groups
respectively, where $k=3$ or $k=9$. The symmetry group of the
SIC~POVM within the Clifford group $\bar{\mathrm{C}}^{(j)}(p)$ is
$\bar{G}_{\mathrm{sym}}^{(j)}=\bar{G}_{\mathrm{sym}}\cap
\bar{\mathrm{C}}^{(j)}(p)$. According to the previous discussions,
at least one of them is identical with $\bar{G}_{\mathrm{sym}}$, and
each $\bar{G}_{\mathrm{sym}}^{(j)}$ different from
$\bar{G}_{\mathrm{sym}}$ is a $p$-group of order $27$, which is
isomorphic to the Sylow $p$-subgroup $\bar{P}_1$ of the standard
Clifford group, and thus contains three HW groups (see
\sref{sec:HWs}). Moreover, each $\bar{G}_{\mathrm{sym}}^{(j)}$
contains at least one HW group whose associated Clifford group
contains $\bar{G}_{\mathrm{sym}}$. In virtue of this observation, we
can easily determine the symmetry group of any group covariant
SIC~POVM in dimension three, no matter whether it is  a subgroup of
the standard Clifford group.

If $|\bar{G}_{\mathrm{sym}}|=27$, $\bar{G}_{\mathrm{sym}}$  is  the
intersection of the three Clifford groups associated with the three
HW groups contained in $\bar{G}_{\mathrm{sym}}$ respectively; in
other words, $\bar{G}_{\mathrm{sym}}^{(1)},
\bar{G}_{\mathrm{sym}}^{(2)}, \bar{G}_{\mathrm{sym}}^{(3)}$ all
coincides with $\bar{G}_{\mathrm{sym}}$. Hence, starting from any HW
group contained in $\bar{G}_{\mathrm{sym}}$, the symmetry group
within its Clifford group is the same. If
$|\bar{G}_{\mathrm{sym}}|=54$, $\bar{G}_{\mathrm{sym}}$ also
contains three HW groups; however, only one of the three groups
$\bar{G}_{\mathrm{sym}}^{(1)}, \bar{G}_{\mathrm{sym}}^{(2)},
\bar{G}_{\mathrm{sym}}^{(3)}$ is identical with
$\bar{G}_{\mathrm{sym}}$. That is, $\bar{G}_{\mathrm{sym}}$ is a
subgroup of only one of the Clifford groups associated with the
three HW groups  respectively. If $|\bar{G}_{\mathrm{sym}}|=216$,
$\bar{G}_{\mathrm{sym}}$ contains nine HW groups, and   is also a
subgroup of only one of the Clifford groups of these HW groups
respectively. In either of the later two cases, starting from
different HW groups, we may``see'' different symmetry groups, if we
only consider symmetry operations within the Clifford group of the
given HW group.

Now it is straightforward to extend the above analysis to  show that
the extended symmetry group $\overline{EG}_{\mathrm{sym}}$ of a
group covariant SIC~POVM is a subgroup of some extended Clifford
group. Suppose the extended symmetry group of the SIC~POVM contains
antiunitary operations (otherwise the claim is already proved); then
$\bar{G}_{\mathrm{sym}}$ is a normal subgroup of
$\overline{EG}_{\mathrm{sym}}$ with index 2. If
$|\bar{G}_{\mathrm{sym}}|=27$, $\bar{G}_{\mathrm{sym}}$ contains
three HW groups all of which are  normal. At least one of the three
HW groups is also normal in $\overline{EG}_{\mathrm{sym}}$, and
$\overline{EG}_{\mathrm{sym}}$ is a subgroup of the extended
Clifford group of this HW group. If $|\bar{G}_{\mathrm{sym}}|=54$,
$\bar{G}_{\mathrm{sym}}$ contains three HW groups one of which is
normal, and the other two are conjugated to each other. The normal
HW group in $\bar{G}_{\mathrm{sym}}$ must remain normal in
$\overline{EG}_{\mathrm{sym}}$, otherwise all three HW groups would
be conjugated to each other in $\overline{EG}_{\mathrm{sym}}$, which
contradicts the fact that the index of $\bar{G}_{\mathrm{sym}}$ in
$\overline{EG}_{\mathrm{sym}}$ is 2. The same analysis is also
applicable  when $|\bar{G}_{\mathrm{sym}}|=216$. So we obtain

\begin{theorem}
\label{thm:main2}In dimension three, each group covariant SIC~POVM
may be covariant with respect to three or nine HW groups, and its
symmetry group within the standard  Clifford group contains at least
three HW  groups. Furthermore, the (extended) symmetry group of the
SIC~POVM is a subgroup of at least one of the (extended) Clifford
groups associated with these HW groups respectively.
\end{theorem}

In dimension three, not surprisingly, there are counterexamples to
corollary~\ref{cor:generalEqui} in \sref{sec:general}, since a
SIC~POVM may be covariant under more than one HW groups. If a
unitary operation maps an HW covariant SIC~POVM to another HW
covariant one, then it must map one of the HW groups
$D^{(1)},\ldots,D^{(k)}$ to the standard HW group $D$. Let $U^{(j)}$
be a unitary transformation that maps $D^{(j)}$ to $D$. By applying
two transformations $U^{(j)}, U^{(l)}$ with $1\leq j,l \leq k$
respectively to a given HW covariant SIC~POVM, two other HW
covariant SIC~POVMs are obtained. These two SIC~POVMs  are on the
same orbit of the standard Clifford group if and only if $D^{(j)},
D^{(l)}$ are conjugated to each other in $G_{\mathrm{sym}}$. Hence,
for each HW covariant SIC~POVM  in dimension three, the number of
 orbits of unitarily equivalent SIC~POVMs is equal to the number of
conjugacy classes of the HW groups contained in the symmetry group
$G_{\mathrm{sym}}$.

In addition, two SIC~POVMs are on the same orbit of the Clifford
group if and only if they are on the same orbit of the extended
Clifford group. To demonstrate this point, without loss of
generality, we may assume that  $G_{\mathrm{sym}}$ contains the
Sylow $p$-subgroup $P_1$ of the Clifford group. Then each fiducial
vector of the SIC~POVM  can be written in the form
$\frac{1}{\sqrt{2}}(1,\rme^{\rmi t},0)$ up to permutations of the
three entries. Since the symmetry group of any SIC~POVM in this
family contains antiunitary operations, the orbit length in the
Clifford group is the same as that in the extended Clifford group
\cite{App05} (see also \sref{sec:sicpom3}). So we obtain
\begin{corollary}\label{cor:3Equi}
In dimension three, for each   SIC~POVM covariant with respect to
the HW group, there are three orbits (both under the Clifford group
and the extended Clifford group) of equivalent SIC~POVMs  if its
symmetry group has order 27, and two orbits  if the symmetry group
has order 54 or 216. In either case, the orbits of equivalent
SIC~POVMs  are connected to each other by unitary transformations
that map additional HW groups contained in the symmetry group within
the standard Clifford group to the standard HW group.
\end{corollary}

\section{\label{sec:sicpom3}SIC~POVMs in three-dimensional Hilbert space}
SIC~POVMs in dimension three are special in many aspects. A SIC~POVM
may be covariant with respect to more than one HW groups; SIC~POVMs
on different orbits may be equivalent even if their fiducial vectors
have  stability groups (within the standard Clifford group) of
different orders; in particular, there exist continuous orbits of
fiducial vectors. In this section, we establish a complete
equivalence relation of SIC~POVMs  on different orbits in dimension
three, and classify inequivalent ones according to the geometric
phases associated with fiducial vectors. In addition, we show that
additional SIC~POVMs can be constructed by regrouping of the
fiducial vectors from different SIC~POVMs which may or may not be on
the same orbit of the extended Clifford group. The  methods used for
SIC~POVMs in dimension three are also applicable to SIC~POVMs in
other dimensions, no matter prime or not.

\subsection{\label{sec:infinite}Infinitely many inequivalent
SIC~POVMs}

There is a one-parameter family of  fiducial vectors in dimension
three,
\begin{eqnarray}
\label{eq:fiducial3}
 |\psi_\mathrm{f}(t)\rangle=\frac{1}{\sqrt{2}}
\left(
  \begin{array}{c}
    0 \\
    1 \\
   - \mathrm{e}^{\rmi t} \\
  \end{array}
\right),
\end{eqnarray}
and for each distinct orbit, there is exactly one value of
$t\in[0,\frac{\pi}{3}]$, such that $|\psi_\mathrm{f}(t)\rangle$ is
on the orbit. There are three kinds of orbits, two exceptional
orbits corresponding to the endpoints $t=0$ and $t=\frac{\pi}{3}$
respectively, and infinitely many generic orbits corresponding to
$0<t<\frac{\pi}{3}$ \cite{Zau99, RBSC04, App05}.

According to Appleby \cite{App05},  the order of the stability group
within the Clifford group (extended
 Clifford group) of each fiducial vector is 24, 6, 3 (48,
12, 6)  for the three kinds of orbits respectively. Hence, the
number of SIC~POVMs on each orbit is 1, 4, 8 respectively. The orbit
length is the same under both the Clifford group and the extended
Clifford group. The stability group (within the extended Clifford
group) of the exceptional vector $|\psi_\mathrm{f}(0)\rangle$
consists of all operations of the form $[F,\bm{0}]$ with $F\in
\mathrm{ESL}(2,3)$. The stability group of the exceptional vector
$|\psi_\mathrm{f}(\frac{\pi}{3})\rangle$ is generated by the unitary
operation
\begin{eqnarray}
&[F,\chi]=\left[\left(%
\begin{array}{cc}
  -1 & 0 \\
  -1 & -1 \\
\end{array}%
\right) ,\left(%
\begin{array}{c}
  0 \\
  1 \\
\end{array}%
\right) \right]&
\end{eqnarray}
and antiunitary operation
\begin{eqnarray}
&{[A,\chi]}=\left[\left(%
\begin{array}{cc}
  1 & 0 \\
  0 & -1 \\
\end{array}%
\right) ,\left(%
\begin{array}{c}
  0 \\
  1 \\
\end{array}%
\right) \right].&
\end{eqnarray}
For a generic vector $|\psi_{\mathrm{f}}(t)\rangle$ with
$0<t<\frac{\pi}{3}$, the stability group is generated by the unitary
operation
\begin{eqnarray}
\label{staGeneric1}
&[F,\chi]^2=\left[\left(%
\begin{array}{cc}
  1 & 0 \\
  -1 & 1 \\
\end{array}%
\right) ,\left(%
\begin{array}{c}
  0 \\
  0 \\
\end{array}%
\right) \right]&
\end{eqnarray}
and antiunitary operation
\begin{eqnarray}
\label{staGeneric2}
&{[F,\chi]\circ[A,\chi]}=\left[\left(%
\begin{array}{cc}
  -1 & 0 \\
  -1 & 1 \\
\end{array}%
\right) ,\left(%
\begin{array}{c}
  0 \\
  0 \\
\end{array}%
\right) \right].&
\end{eqnarray}
For each generic vector, the stability group is independent of $t$;
in particular, the two fiducial vectors
$\hat{J}|\psi_{\mathrm{f}}(t)\rangle=|\psi_{\mathrm{f}}(-t)\rangle$
and $|\psi_{\mathrm{f}}(t)\rangle$ have the same stability group,
where $\hat{J}$ is the complex conjugation operator (see
\sref{sec:Clifford}).

For each generic orbit, the eight SIC~POVMs on the orbit form four
pairs $[F_k,\bm{0}]|\psi_{\mathrm{f}}(t)\rangle,
[F_k,\bm{0}]\hat{J}|\psi_{\mathrm{f}}(t)\rangle$ (here each fiducial
vector  represents the SIC~POVM containing it) for $k=1,2,3,4$,
where
\begin{eqnarray}\label{eq:pairs}
\fl F_1=\left(
      \begin{array}{cc}
        1 & 0 \\
        0 & 1 \\
      \end{array}
    \right),\qquad
    F_2=\left(
      \begin{array}{cc}
        0& 1 \\
        2 & 0 \\
      \end{array}
    \right),\qquad
    F_3=\left(
      \begin{array}{cc}
        0& 1 \\
        2 & 1 \\
      \end{array}
    \right),\qquad
        F_4=\left(
      \begin{array}{cc}
        0& 1 \\
        2 & 2 \\
      \end{array}
    \right).
\end{eqnarray}
 Within the Clifford group, the two SIC~POVMs in
each pair share the same symmetry group, which is a Sylow 3-subgroup
of the Clifford group. There is a one-to-one correspondence between
the four pairs of SIC~POVMs and the four Sylow 3-subgroups of the
Clifford group. The symmetry group of each SIC~POVM contains three
HW groups; all of which are normal subgroups. The intersection of
any two symmetry groups of two SIC~POVMs from different pairs
respectively is exactly the standard HW group. For the exceptional
orbit corresponding to $t=\frac{\pi}{3}$, each SIC~POVM is invariant
under the complex conjugation operation, so the two SIC~POVMs in
each pair merge to one. The symmetry group of each SIC~POVM is the
normalizer (within the Clifford group) of a Sylow 3-subgroup, and
contains three HW groups too. However, only the standard HW group is
a normal subgroup, and the other two are conjugated to each other.
The intersection of any two symmetry groups is the group generated
by $[-\bm{1},\bm{0}]$ and the standard HW group. For the exceptional
orbit corresponding $t=0$, there is only one SIC~POVM, and its
symmetry group is the Clifford group, which contains nine HW groups.
The standard HW group is a normal subgroup, while the other eight HW
groups are conjugated to each other.

We are now ready to show the equivalence relation of SIC~POVMs among
different orbits in virtue of theorem~\ref{thm:main2} and
corollary~\ref{cor:3Equi} derived in \sref{sec:special}. Note that
for each $t\in[0,\frac{\pi}{3}]$, the symmetry group of the SIC~POVM
generated from the fiducial vector $|\psi_{\mathrm{f}}(t)\rangle$
contains as a subgroup the Sylow 3-subgroup $P_1$ of the Clifford
group. According to \eref{eq:HWpermute2} and \eref{eq:HWpermute3},
$U=\mathrm{diag}(1,\mathrm{e}^{-2\rmi\pi/9},\mathrm{e}^{-4\rmi\pi/9})$
is a unitary transformation that permutes the three HW groups
contained in $P_1$. According to corollary~\ref{cor:3Equi}, the
SIC~POVMs on the three orbits generated from
$|\psi_\mathrm{f}(t)\rangle$, $U^\dag|\psi_\mathrm{f}(t)\rangle$ and
$U^{\dag2}|\psi_\mathrm{f}(t)\rangle$  respectively are unitarily
equivalent. That is, SIC~POVMs on the three orbits corresponding to
the parameters $t, \frac{2\pi}{9}+t, \frac{2\pi}{9}-t$ respectively
for each $t\in [0, \frac{\pi}{9}]$ (when $t=0$ or $t=\frac{\pi}{9}$,
two of the three orbits may merge) are unitarily equivalent.
Moreover, SIC~POVMs on any two different orbits corresponding to
$t\in[0,\frac{\pi}{9}]$ are not equivalent. Hence, there are two
orbits of equivalent SIC~POVMs  for each exceptional orbit, and
three orbits of equivalent SIC~POVMs for each generic orbit with
$t\neq\frac{\pi}{9}, \frac{2\pi}{9}$.

The equivalence of the exceptional orbit with $t=0$ and the generic
orbit with $t=\frac{2\pi}{9}$ is particularly surprising at first
glance, since they have stability groups of different orders (within
the standard Clifford group). Equally surprising is the equivalence
of the exceptional orbit with $t=\frac{\pi}{3}$ and the generic
orbit with $t=\frac{\pi}{9}$.

In addition, the (extended) symmetry group of any SIC~POVM except
those on the orbit with $t=\frac{\pi}{9}$ or $t=\frac{2\pi}{9}$   is
a subgroup of the standard (extended) Clifford group. For each
SIC~POVM on the orbit with $t=\frac{\pi}{9}$ or $t=\frac{2\pi}{9}$,
its (extended) symmetry group is a subgroup of the (extended)
Clifford group associated with another HW group contained in the
symmetry group within the standard Clifford group.

In conjunction with theorem~\ref{thm:main1}, we obtain a quite
surprising conclusion: Among all (HW) group covariant SIC~POVMs in
prime dimensions, the SIC~POVMs in dimension three on the orbits
generated from the fiducial vectors in \eref{eq:fiducial3} with
$t=\frac{\pi}{9}$ and  $t=\frac{2\pi}{9}$ respectively are the only
ones whose (extended) symmetry groups  are not subgroups of the
standard (extended) Clifford group.

If we denote by $\mathrm{A}_{\pm}$,  $\mathrm{B}_{\pm}$,
$\mathrm{C}_{\pm}$ the six  SIC~POVMs (two on each of the three
orbits of equivalent SIC~POVMs) containing the fiducial vectors
$|\psi_{\mathrm{f}}(\pm t)\rangle$,
$|\psi_{\mathrm{f}}(\pm(\frac{2\pi}{9}-t))\rangle$,
$|\psi_{\mathrm{f}}(\pm(\frac{2\pi}{9}+t))\rangle$ respectively,
then the transformation among the six SIC~POVMs induced by
$U^{\dag}$ can be illustrated as follows:
\begin{eqnarray}
\mathrm{A}_{+}\rightarrow\mathrm{C}_{+}\rightarrow \mathrm{B}_{-},
\qquad \mathrm{A}_{-}\rightarrow
\mathrm{B}_{+}\rightarrow\mathrm{C}_{-}.
\end{eqnarray}
Interestingly, the three SIC~POVMs $\mathrm{A}_+,
\mathrm{B}_-,\mathrm{C}_+$ cycles among the three obits in the
opposite direction compared with the other three SIC~POVMs
$\mathrm{A}_-, \mathrm{B}_+ , \mathrm{C}_- $. Although the SIC~POVMs
on the three orbits with $t, \frac{2\pi}{9}-t, \frac{2\pi}{9}+t$
respectively are equivalent, the orbits themselves are not
equivalent in the sense that there is no unitary or antiunitary
transformation that can map all SIC~POVMs or fiducial vectors on one
of the three orbits to that on another one. For example, under the
transformation induced by $U^{\dag}$, only six out of the 24
SIC~POVMs on the three orbits are permuted among each other; the
other 18 SIC~POVMs are no longer on any of the three orbits. This
point will become more clear when we study the additional SIC~POVMs
constructed by regrouping of the fiducial vectors in
\sref{sec:regrouping}.

To better characterize those inequivalent SIC~POVMs, we need to find
some invariants that can distinguish them.  The simplest invariant
involves three different  vectors in a SIC~POVM. Let
$\rho_j=|\psi_j\rangle\langle\psi_j|$ for $j=1,2,3$, where
$|\psi_j\rangle$s are three different vectors in a SIC~POVM. The
trace of the triple product $\mathrm{tr}(\rho_1\rho_2\rho_3)$ is
invariant under unitary transformation. This invariant has been
applied by Appleby {\it et al.} \cite{ADF07} to studying the set of
occurring probabilities of measurement outcomes of SIC~POVMs.
According to \eref{eq:SIC},
$|\mathrm{tr}(\rho_1\rho_2\rho_3)|=\frac{1}{8}$ for $d=3$, so the
relevant invariant is the phase  of the trace,
$\phi^\prime=\mathrm{arg}[\mathrm{tr}(\rho_1\rho_2\rho_3)]$, with
$-\pi\leq\phi^\prime<\pi$. Since odd permutations or complex
conjugation of the three states reverses the sign of the phase, we
shall be concerned with the absolute value of the phase,
$\phi=|\mathrm{arg}[\mathrm{tr}(\rho_1\rho_2\rho_3)]|$, with
$0\leq\phi\leq \pi$. Now  $\phi$ is independent of the permutations
and complex conjugation of the three states.  Recall that the phase
$\phi$  defined above is exactly the discrete  geometric phase
associated with the three  vectors $|\psi_1\rangle, |\psi_2\rangle,
|\psi_3\rangle$, which is the additional phase appearing after
traversing the geodesic triangle with the three vectors as vertices
in the projective Hilbert space  \cite{Ber84, AA87}. It is also
known as the Bargmann invariant \cite{Bar64}. Another invariant
associated with the three vectors is the set of eigenvalues of the
sum $M=\rho_1+\rho_2+\rho_3$. However, some simple algebra shows
that it is not an independent invariant:
\begin{eqnarray}
&&\mathrm{tr}\bigl(\rho_1+\rho_2+\rho_3\bigr)=3,\nonumber\\
&&\mathrm{tr}\bigl[\bigl(\rho_1+\rho_2+\rho_3\bigr)^2\bigr]
 =\frac{9}{2},\nonumber\\
&&\mathrm{tr}\bigl[\bigl(\rho_1+\rho_2+\rho_3\bigr)^3\bigr]
=\frac{15}{2}+\frac{3}{4}\cos\phi.
\end{eqnarray}
Since the eigenvalues of a $3\times3$ matrix are determined  by its
lowest three moments, the eigenvalues of $M$ are determined by
$\phi$. Thus $\phi$ is the only independent invariant associated
with the three  vectors.

\begin{figure}
\centering
  \includegraphics[width=10cm]{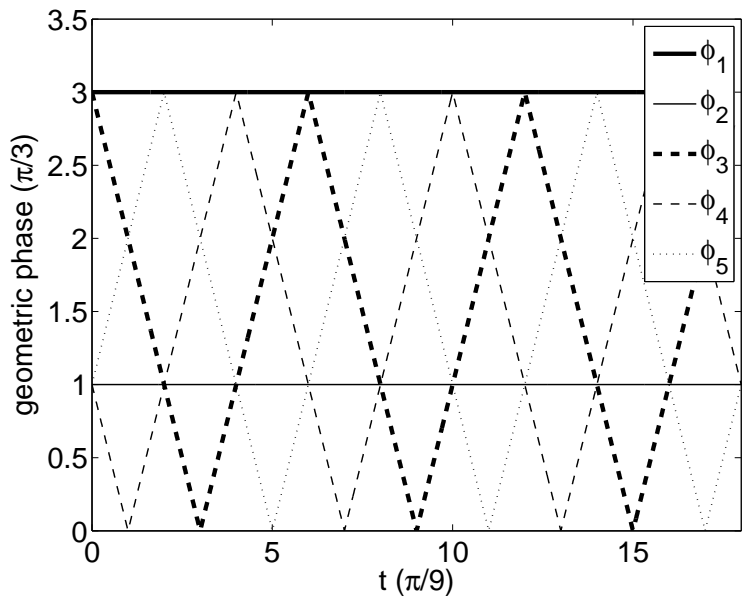}\\
  \caption{\label{fig:phase}  Geometric phases associated with five different triples of vectors respectively in the
  SIC~POVM generated from the fiducial vector in \eref{eq:fiducial3} for $t\in[0,2\pi]$.
  See the main text and \tref{tab:phase} for the meaning of the five different phases.}
\end{figure}

Given a SIC~POVM generated from the fiducial vector
$|\psi_\mathrm{f}(t)\rangle$ in \eref{eq:fiducial3}, due to group
covariance, without loss of generality, we may assume
$\rho_1=\rho_f=|\psi_\mathrm{f}(t)\rangle\langle\psi_\mathrm{f}(t)|$.
There are ${8 \choose 2}=28$ different choices for the remaining two
 vectors, $|\psi_2\rangle, |\psi_3\rangle$.  However, some analysis of the
symmetry group of the SIC~POVM reveals that $\phi$ may take at most
five different values. \Tref{tab:phase} shows the five distinct
geometric phases associated with five different triples of vectors
in the SIC~POVM on the orbit with $t\in [0, \frac{\pi}{9}]$. Figure
\ref{fig:phase} shows the variation of the five phases with $t$ in a
wider range. The two phases $\phi_1, \phi_2$ are independent of the
parameter $t$. The other three phases $\phi_3, \phi_4, \phi_5$ are
periodic functions of $t$ with the same shape and  period
$\frac{2\pi}{3}$, but are shifted from each other by
$\pm\frac{2\pi}{9}$. If we do not distinguish the three phases
$\phi_3, \phi_4, \phi_5$, then the pattern displays a period of
$\frac{2\pi}{9}$, with an additional mirror symmetry about
$t=\frac{k\pi}{9}$ for $k=0, \pm1, \pm2,\ldots$. It is clear from
the figure that any two SIC~POVMs on two different orbits
respectively with $t\in[0,\frac{\pi}{9}]$  are not equivalent. By
contrast, the equivalence of SIC~POVMs on the three orbits
corresponding to $t, \frac{2\pi}{9}-t, \frac{2\pi}{9}+t$
respectively is underpinned.
\begin{table}
\centering
  \caption{\label{tab:phase}  Geometric phases $\phi=|\mathrm{arg}[\mathrm{tr}(\rho_1\rho_2\rho_3)]|$
  associated with five different triples of vectors respectively  of
  the
  SIC~POVM generated from the fiducial vector in \eref{eq:fiducial3}
  for  $t\in[0,\frac{\pi}{9}]$, where
  $\rho_j=|\psi_j\rangle\langle\psi_j|$ for $j=1, 2, 3$, and
  $|\psi_j\rangle$s are three different vectors in the SIC~POVM.
  Here $[Z]$ represents the vector $Z|\psi_{\mathrm{f}}(t)\rangle$, similarly for $[X]$
  etc. Due to group covariance, $|\psi_{\mathrm{f}}(t)\rangle$ is chosen as
  $|\psi_1\rangle$.
  There are 28 different choices in total for the pair $|\psi_2\rangle,
  |\psi_3\rangle$.
  The second column shows the numbers of choices that lead to the specific geometric phases given in the third column.}
\begin{math}
  \begin{array}{ccc}
    \br
    \{|\psi_1\rangle, |\psi_2\rangle, |\psi_3\rangle\} & \textrm{multiplicity} & \textrm{geometric phase}\\
     \hline
    \{[I], [Z], [Z^2]\} & 1 & \phi_1=\pi \\
     \{[I],[X], [Z]\} & 18 & \phi_2=\frac{\pi}{3} \\
    \{[I], [X], [X^2]\} & 3 & \phi_3=\pi-3t \\
   \{[I], [X], [X^2Z]\} & 3 & \phi_4= \frac{\pi}{3}-3t\\
    \{[I], [X], [X^2Z^2]\} & 3 & \phi_5=\frac{\pi}{3}+3t \\
    \br
  \end{array}
  \end{math}
\end{table}

Let $\phi_{\mathrm{min}}$ be the minimum of the five phases listed
in \tref{tab:phase}; then $0\leq\phi_{\mathrm{min}}\leq
\frac{\pi}{3}$, and there is a one-to-one correspondence between
$\phi_{\mathrm{min}}$ and $t$ within the interval
$[0,\frac{\pi}{9}]$, which reads
\begin{eqnarray}
\phi_{\mathrm{min}}=\frac{\pi}{3}-3t.
 \end{eqnarray}
Thus $\phi_{\mathrm{min}}$ uniquely specifies the equivalence class
of a group covariant SIC~POVM in dimension three. Unlike the
parameter $t$, the phases $\phi_j$s  and $\phi_{\mathrm{min}}$ are
intrinsic quantities of the SIC~POVM, which are independent of  the
parametrization. They are especially useful when the SIC~POVM is not
constructed from a fiducial vector or the information about the
symmetry group is missing, such as in the case of  ``hidden''
SIC~POVMs to be studied in \sref{sec:regrouping}.

From the values of the geometric phases in \tref{tab:phase} and
\fref{fig:phase}, it is straightforward to  construct an alternative
proof that, for each SIC~POVM on the orbit with
$t\in[0,\frac{\pi}{3}]$ and $t\neq\frac{\pi}{9},\frac{2\pi}{9}$,
there are no additional unitary or antiunitary symmetry operations
except those already contained in the standard extended Clifford
group. The reasoning is based on the simple fact that these phases
must be preserved under unitary or antiunitary operations.
Interestingly, unlike the case in dimension two, it is not always
possible to map three vectors to another three  vectors in the same
SIC~POVM with unitary or antiunitary operations. In addition, in
quantum state tomography with a SIC~POVM in dimension three, the set
of occurring probabilities is not permutation invariant,
 according to \cite{ADF07}.

As far as tomographic efficiency is concerned, each SIC~POVM is
almost equally good \cite{Sco06, ZTE10}. Although there are
infinitely many inequivalent SIC~POVMs in dimension three, and the
set of occurring probabilities in tomography may depend on the
SIC~POVM chosen. It is interesting to know whether there is any
other application, such that inequivalent SIC~POVMs may lead to a
drastic difference.

\subsection{\label{sec:regrouping}Uncover additional SIC~POVMs by
regrouping of fiducial vectors}

Almost all known SIC~POVMs are constructed from fiducial vectors
under the action of the HW group. In this section, we uncover
additional SIC~POVMs in dimension three by regrouping of the
fiducial vectors.

In addition to the SIC~POVMs generated from the fiducial vectors in
\eref{eq:fiducial3}, there are some ``hidden'' SIC~POVMs composed of
vectors from different orbits or from different SIC~POVMs on the
same orbit. For example, the following nine vectors
$X^j|\psi_{\mathrm{f}}(t_j)\rangle,
ZX^j|\psi_{\mathrm{f}}(t_j)\rangle,
Z^2X^j|\psi_{\mathrm{f}}(t_j)\rangle$ for $j=0,1,2$ with $t_j\in
[0,2\pi)$ also form a SIC~POVM. Although this SIC~POVM is not
constructed from a fiducial vector with the HW group, it is
equivalent to a SIC~POVM on the orbit with
$t=\frac{t_0+t_1+t_2}{3}$, under the unitary transformation
$\mathrm{diag}\bigl(1,\rme^{\rmi(t-t_2)},
\rme^{\rmi(2t-t_0-t_2)}\bigr)$.

Now suppose the eight SIC~POVMs on each generic orbit are divided
into four pairs as  in  \sref{sec:infinite} (see \eref{eq:pairs}).
In  each pair of SIC~POVMs, we can construct six additional
SIC~POVMs by regrouping of the 18 fiducial vectors. For example,
given  $k_1=0, 1, 2$, the following nine vectors
\begin{eqnarray}
\fl Z^{k_2}X^{k_1}|\psi_{\mathrm{f}}(t)\rangle,\qquad
Z^{k_2}X^{k_1+1}|\psi_{\mathrm{f}}(t)\rangle,\qquad
Z^{k_2}X^{k_1+2}|\psi_{\mathrm{f}}(-t)\rangle \qquad\mbox{for}\quad
k_2=0,1,2
\end{eqnarray}
in the first pair of SIC~POVMs form a SIC~POVM, which is unitarily
equivalent to any SIC~POVM on the orbit with $t^\prime=\frac{t}{3}$.
Similarly,  the following nine vectors
\begin{eqnarray}
\fl Z^{k_2}X^{k_1}|\psi_{\mathrm{f}}(t)\rangle,\qquad
Z^{k_2}X^{k_1+1}|\psi_{\mathrm{f}}(-t)\rangle,\qquad
Z^{k_2}X^{k_1+2}|\psi_{\mathrm{f}}(-t)\rangle \qquad\mbox{for}\quad
 k_2=0,1,2
\end{eqnarray}
also form a SIC~POVM, which is unitarily equivalent to any SIC~POVM
on the orbit with $t^{\prime\prime}=-\frac{t}{3}$, and is thus also
unitarily equivalent to any SIC~POVM on the orbit with
$t^\prime=\frac{t}{3}$. By the same token, six additional SIC~POVMs
can be obtained by regrouping of the 18 fiducial vectors  from any
other pair of  SIC~POVMs. Further analysis shows that these 24
additional SIC~POVMs exhaust all SIC~POVMs that can be obtained by
regrouping of the 72 fiducial vectors on each generic orbit. All the
24 additional SIC~POVMs are unitarily equivalent, however, they are
not unitarily (even antiunitarily) equivalent to the original eight
SIC~POVMs. For the exceptional orbit with $t=\frac{\pi}{3}$, no
SIC~POVMs can be obtained by regrouping of the fiducial vectors in
the four SIC~POVMs.

Although the SIC~POVMs on the three orbits with $t,
\frac{2\pi}{9}-t, t+\frac{2\pi}{9}$ respectively for $t\in
(0,\frac{\pi}{9})$ are unitarily equivalent (see
\sref{sec:infinite}), the additional SIC~POVMs obtained by
regrouping of the fiducial vectors for the three orbits respectively
are not unitarily (even antiunitarily) equivalent. This implies in
particular that there is no unitary or antiunitary transformation
that can map all SIC~POVMs on one of the three orbits to that on
another one.

SIC~POVMs obtained by regrouping of fiducial vectors have been known
for dimension four \cite{Gra08a, ZET}. For other dimensions, as far
as the SIC~POVMs found by Scott and Grassl \cite{SG10} are
concerned, such additional SIC~POVMs can be obtained only for the
orbits 8b and 12b (according to the labeling Scheme of Scott and
Grassl) \cite{ZET}. The peculiarity of SIC~POVMs on these orbits is
still a mystery.

\section{\label{sec:con}   Summary}
The equivalence relation of SIC~POVMs on different orbits of the
(extended) Clifford group has been an elusive question in the
community. So is the closely related question: Is the  (extended)
symmetry group of an HW covariant SIC~POVM a subgroup of the
(extended) Clifford group? In this paper we resolve these open
questions for all prime dimensions. More specifically, we prove
that, in any prime dimension not equal to three, each group
covariant SIC~POVM is covariant with respect to a unique HW group;
its (extended) symmetry group is a subgroup of the (extended)
Clifford group. Hence, SIC~POVMs on different orbits are not
equivalent. In dimension three, each group covariant SIC~POVM may be
covariant with respect to three or nine HW groups; its symmetry
group is a subgroup of at least one of the Clifford groups
associated with these HW groups respectively. There may exist two or
three orbits of equivalent SIC~POVMs depending on the order of the
symmetry group.

In addition, we establish a complete equivalence relation among
group covariant SIC~POVMs in dimension three, and classify
inequivalent ones according to the geometric phases associated with
fiducial vectors. Also, we uncover additional SIC~POVMs by
regrouping of the fiducial vectors from different SIC~POVMs which
may or may not be on the same orbit of the extended Clifford group.
The picture of the SIC~POVMs in dimension three is now more
complete. The methods employed are also applicable to SIC~POVMs in
other dimensions.

Our results are an important step towards understanding the
structure of SIC~POVMs in prime dimensions. It would be highly
desirable to extend these results to other dimensions.

\section*{Acknowledgements}
The author  is grateful to Markus Grassl for stimulating discussions
and valuable comments and suggestions on the manuscript, and to
Berthold-Georg Englert and Andrew Scott  for stimulating
discussions. The Centre for Quantum Technologies is funded by the
Singapore Ministry of Education and the National Research Foundation
as part of the Research Centres of Excellence programme.


\section*{References}
\begin{thebibliography}{99}

\bibitem{Zau99}
Zauner G 1999 Ph.D. thesis, University of Vienna. Available online
at \url{http://www.mat.univie.ac.at/~neum/papers/physpapers.html}

\bibitem{RBSC04}
Renes J M, Blume-Kohout R, Scott A J and Caves C M 2004 {\it J.
Math.\ Phys.\ } \textbf{45} 2171


\bibitem{App05}
Appleby D M 2005 {\it J.\ Math.\ Phys.\ } \textbf{46} 052107

\bibitem{SG10} Scott A J and  Grassl M 2010
{\it J.\ Math.\ Phys.\ } \textbf{51} 042203 (arXiv:0910.5784
[quant-ph])

\bibitem{Fuc02} Fuchs C A 2002   arXiv:quant-ph/0205039v1

\bibitem{Sco06}
Scott A J 2006 {\it J. Phys.\ A: Math.\ Gen.\ } \textbf{39} 13507


\bibitem{ZTE10}
Zhu H, Teo Y S  and  Englert B-G, Quantum State Tomography with
Joint SIC~POMs and Product SIC~POMs (in preparation)



\bibitem{DGS75}  Delsarte P,  Goethals J M and Seidel J J 1975
{\it Philips Res. Rep.} {\bf 30} 91

\bibitem{Gra04} Grassl M 2004 {\it Proc.  2004 ERATO Conf.
on Quantum Information Science} (Tokyo, September, 2004) pp 60--61
(arXiv:quant-ph/0406175)

\bibitem{Hog98} Hoggar S G 1998 {\it Geom. Dedicata}  {\bf 69} 287

\bibitem{Gra05}  Grassl M 2005 {\it Electron. Notes  Discrete Math.} {\bf 20} 151


\bibitem{Gra06}  Grassl M 2006 Finding equiangular lines in complex
space {\it MAGMA 2006 Conference} (Technische Universit\"at Berlin,
July, 2006) available online at
\url{http://magma.maths.usyd.edu.au/Magma2006/}

\bibitem{Gra08a} Grassl M 2008 Seeking symmetries of SIC-POVMs
{\it Seeking SICs: A Workshop on Quantum Frames and Designs}
(Perimeter Institute, Waterloo, October, 2008) available online at
\url{http://pirsa.org/08100069/}

\bibitem{Gra08b} Grassl M 2008 {\it
  Lect. Notes Comput. Sci.} {\bf 5393} 89


\bibitem{Woo04}Wootters  W K 2004   arXiv:quant-ph/0406032

\bibitem{ADF07} Appleby D M, Dang H  and  Fuchs C  2007
arXiv:0707.2071 [quant-ph]

\bibitem{App08} Appleby D M 2009 AIP
  Conf. Proc. {\bf 1101} 223

\bibitem{LS73}  Lemmens P W H and Seidel J J 1973  {\it  J. Algebra} {\bf 24} 494

\bibitem{AFF09}   Appleby D M,   Flammia S T  and Fuchs C 2009
arXiv:1001.0004 [quant-ph]


\bibitem{App09} Appleby D M 2009   arXiv:0909.5233 [quant-ph]

\bibitem{Ber84} Berry  M V 1984 {\it Proc. R. Soc. A} {\bf 392} 45


\bibitem{AA87}  Aharonov Y and Anandan J 1987 \PRL {\bf 58} 1593


\bibitem{Bar64} Bargmann V  1964 {\it J. Math. Phys.} {\bf 5} 862

\bibitem{Bli917}  Blichfeldt H F 1917 {\it Finite Collineation Groups \/}
(Chicago, IL: University of Chicago Press)

\bibitem{Dic58} Dickson L  E 1958 {\it Linear Groups, with an Exposition of
the Galois Field Theory\/} (New York: Dover)

\bibitem{KS04} Kurzweil H and  Stellmacher B 2004 {\it The Theory of Finite
Groups, An Introduction\/} (New York: Springer)


\bibitem{Hum75} Humphreys J  E 1975 {\it Amer. Math. Monthly} {\bf 82} 21


\bibitem{ABC07}  Appleby D M,  Bengtsson I and Chaturvedi S 2007   arXiv:0710.3013 [quant-ph]

\bibitem{Fla06} Flammia S T  2006 {\it J. Phys.\ A: Math.\ Gen.\ } {\bf 39}
13483


\bibitem{JL01} James G and Liebeck M 2001 {\it Representations and Characters of
Groups\/} (Cambridge: Cambridge University Press)

\bibitem{CR62} Curtis C W and  Reiner I 1962 {\it Representation Theory
of Finite Groups and Associative Algebras\/} (New York: Wiley)

\bibitem{NC00}Nielsen M A  and Chuang I L 2000 \textit{Quantum Computation and
Quantum Information\/} (Cambridge: Cambridge University Press) p 461


\bibitem{Sib74} Sibley D A 1974 {\it Journal of algebra} {\bf 32} 286

\bibitem{Mac75} Mackiw G  1975 {\it Amer. Math. Monthly} {\bf 82} 64

\bibitem{ZET}
Zhu H, Englert B-G, and Teo Y S (unpublished)






\end{thebibliography}
\end{document}